\documentclass[a4paper,onecolumn,superscriptaddress,amsmath,amssymb,aps,prl,nobibnotes]{revtex4-2}
\usepackage[utf8]{inputenc}
\usepackage[T1]{fontenc}
\usepackage{graphicx}
\graphicspath{ {./figures/} }

\usepackage{amsthm}
\usepackage{float}
\usepackage{eurosym}

\usepackage[normalem]{ulem}

\usepackage{xcolor}
\usepackage{comment}

\begin{document}

\title{Demand-driven design of bicycle infrastructure networks\\ for improved urban bikeability}

\author{Christoph Steinacker}
\affiliation{Chair for Network Dynamics, Center for Advancing Electronics Dresden (cfaed) and Institute for Theoretical Physics, Technische Universit\"at Dresden, 01062 Dresden, Germany}

\author{David-Maximilian Storch}
\affiliation{Chair for Network Dynamics, Center for Advancing Electronics Dresden (cfaed) and Institute for Theoretical Physics, Technische Universit\"at Dresden, 01062 Dresden, Germany}

\author{Marc Timme}
\affiliation{Chair for Network Dynamics, Center for Advancing Electronics Dresden (cfaed) and Institute for Theoretical Physics, Technische Universit\"at Dresden, 01062 Dresden, Germany}

\author{Malte Schr\"oder}\email[Corresponding author: ]{malte.schroeder@tu-dresden.de}
\affiliation{Chair for Network Dynamics, Center for Advancing Electronics Dresden (cfaed) and Institute for Theoretical Physics, Technische Universit\"at Dresden, 01062 Dresden, Germany}

\begin{abstract}
    Cycling is a crucial part of sustainable urban transportation. Promoting cycling critically relies on a sufficiently developed bicycle infrastructure. However, designing efficient bike path networks constitutes a complex problem that requires balancing multiple constraints while still supporting all cycling demand.
    Here, we propose a framework to create families of efficient bike path networks by explicitly taking into account the demand distribution and cyclists' route choices based on safety preferences. 
    By reversing the network formation process and iteratively removing bike paths from an initially complete bike path network and continually updating cyclists' route choices, we create a sequence of networks that is always adapted to the current cycling demand. 
    We illustrate the applicability of this demand-driven planning scheme for two cities. A comparison of the resulting bike path networks with those created for homogenized demand enables us to quantify the importance of the demand distribution for network planning. The proposed framework may thus enable quantitative evaluation of the structure of current and planned bike path networks and support the demand-driven design of efficient infrastructures.
\end{abstract}

\maketitle

Human mobility critically depends on the existing infrastructure underlying it \cite{Banister2008, Mazzoncini2020, Standen2019}. 
The transition to more sustainable mobility in particular requires a sufficiently developed infrastructure to promote, for example, cycling over motorized mobility for short and medium-distance intra-urban trips \cite{Creutzig2016, Creutzig2015, IPCC2014, Li2022_scaling}.
During the \mbox{COVID-19} pandemic, a number of cities such as Paris, New York and Bogotá pushed to open more street space to cyclists, expanded the size of side-walks, or blocked car traffic in order to enable social distancing \cite{Rhoads2021}. Similarly, many cities have vouched to invest into cycling infrastructure \cite{BBC2020, WRI2020, WorldBank2020, Guardian2020} and are gaining increasing support among the population for these investments \cite{Guardian2021}. 

In general, designing suitable and efficient infrastructure networks constitutes an intricate problem as the networks are subject to multiple, often opposing technical, economic and social constraints \cite{Jackson2010_social, Schoreder2018_hysteretic, Aldous2019, Frangopol2007}. Examples for efficient network structures can be found in various biological \cite{Tero2010_rules, Katifori2010_damage, Ronellenfitsch2016_global, Karschau2020_resilience} and social networks \cite{Kleinberg2000_navigation, cohen00_resilience, Molkenthin2018Adhesion}, balancing resource and energy costs with efficient physical or information transport and robustness to failures. For mobility and infrastructure networks, examples of efficient topologies include the core-periphery structure of air travel networks, balancing the cost of direct flights with the inconvenience of transfers \cite{Gastner2006_optimal, Barthelemy2006_optimal, Verma2016_emergence, Cardillo2013_emergence, Tero2010_rules}, and the emergent backbone structure in street networks of cities \cite{Barthelemy2011_spatial, Scellato2006_backbone, Barethelemy2008_modeling, Kirkley2018_betweenness}.

For the design of bike path networks, three major constraints include: 
(i) Budget constraints limit the total length of bike paths, as for most physical networks, e.g. due to construction or maintenance cost \cite{Duthie2014_optimization, Gastner2006_optimal}. 
(ii) Bike path networks have to support the mobility demand, enabling fast travel between frequented locations without large detours \cite{Bil2015, Barthelemy2011_spatial, Gastner2006_optimal, Verma2016_emergence}. 
(iii) In addition to fast travel, bike path networks should enable safe travel of cyclists along highly frequented routes \cite{Munoz2013_topo}. 
Different implementations of bike path infrastructure weigh these constraints differently. For example colored bike lanes on the street do not cost much but add only little safety of cyclists whereas physically separated bike lanes greatly improve cyclists' safety but require more space and investment \cite{RikdeGroot2016_designmanual, Bushell2013_costs}.

From a network structure perspective, each of these aspects is simple to understand individually. For example, a connected bike path network in a city that minimizes the required budget is simply given by the minimum spanning tree of the underlying street network \cite{Barthelemy2006_optimal, Barthelemy2018, Scellato2006_backbone}. More sophisticated approaches for finding connected network structures have recently been proposed based on percolation processes \cite{Natera2020_datadriven, Olmos2020_data, Szell2021_growing} to optimize the connectivity of the bike path network. However, connectivity alone is not sufficient to support the demand since routes along streets equipped with bike paths would likely be indirect and require large detours. Shortest path trees would optimally support the demand only from and to a single location. In contrast, direct routes between other locations would require cycling along streets without a dedicated bike infrastructure and would not be as safe \cite{Menghini2010_routechoice, Broach2012_where}. Finally, the safest and most convenient network for cyclists, where a bike path exists along every street, would naturally exceed any reasonable budget constraints \cite{Duthie2014_optimization, Guerreiro2018_data}. 
However, efficient bike path networks have to simultaneously adhere to all three constraints to enable both convenient and safe travel with feasible investment \cite{Banister2001, Buehler2016, Caulfield2012}.

In this article, we propose a framework for constructing a family of efficient bike path networks, all adapted to the given street network and demand distribution. The algorithm realizes inverted network growth: Based on a simplified cyclist routing model and starting from a network fully equipped with bike paths, the algorithm generates a sequence of bike path networks by successively removing bike paths from the street segments with the least impact given the usage patterns in the current network. We observe that both convenience and safety of cycling in the network remain high, even if only a small part of the street network remains equipped with bike paths. Application of the algorithm to synthetic, homogeneous demand conditions enables us to quantify the importance of the cycling demand for the structure of the resulting bike path networks. The proposed framework is extendable to include different routing models and its applicability to different street networks and demand distributions may support planning of bike path networks that promote a desired cycling demand.

\section*{Results}

\subsection*{Cyclist route choice model}
Route choice and traffic assignment constitute complex problems in modeling the dynamics of human mobility, not least due to congestion and other collective phenomena induced by interactions among individuals. For car traffic, one prominent example of complex collective dynamics is Braess' paradox \cite{Wardrop1952, Youn2008_anarchy} emerging from the selfish routing decision of drivers. For cyclists, Braess paradox and congestion is less relevant due to the smaller spatial footprint compared to cars \cite{Paulsen2019_bikecongestion}. However, cyclists often share infrastructure with cars and their route choice is not only based on the fastest route as often assumed in general transportation or information routing networks \cite{Wardrop1952, Daganzo1977} but also on safety along potential routes. Studies comparing the routes taken by cyclists with potential alternative options consistently find a strong tendency to avoid dangerous traffic situations or inconvenient routes \cite{Menghini2010_routechoice, Broach2012_where}. As cyclists are highly sensitive to traffic volume on streets without separate bike infrastructure, these studies predict that cyclists may prefer significant detours to avoid cycling on busy streets. Other factors that deter cyclists from using a specific route are steep inclines and unsignaled turns, in particular left turns where cyclists may have to wait in the middle of the street \cite{Menghini2010_routechoice, Broach2012_where}. In general, the preference for convenient and safe travel is higher for leisure trips than for commuting cyclists, except for the avoidance of high traffic (possibly due to commuting trips typically taking place under generally busier traffic conditions). Safety thus constitutes a substantial factor in designing bike path networks to promote cycling \cite{McNeil2015, Natera2020_datadriven, Olmos2020_data, Stuelpnagel2022_howsafe}.

We map these route choice preferences to a shortest path problem on a cyclist preference graph $G=(V,E)$ with $N=|V|$ nodes (intersections) and $M=|E|$ edges (street segments). We derive the preference graph $G$ from the physical street network $G^\mathrm{street}$. Both graphs share the same set of nodes $V$. 
Each edge $e_{ij} \in E$ in the cyclist preference graph represents a street segment that connects intersections $i,j \in V$ and is assigned a perceived distance 
\begin{equation}
    l_{ij}=l^\mathrm{street}_{ij}\,p_{ij} \,. \label{eq:weighted_edge_distance}
\end{equation}
Here, $l^\mathrm{street}_{ij}$ denotes the physical length of the corresponding street segment $e^\mathrm{street}_{ij}$ in the street network and $p_{ij} \in \{p^B_{ij}, p^0_{ij}\}$ is the penalty factor summarizing cyclists preferences against riding along a street segment $e_{ij}$. The set of street segments equipped with bike paths, $E_B\subseteq E$, contains street segments $e_{ij} \in E_B$ without distance penalty, $p^B_{ij} = 1$. Street segments not in this set, $e_{ij} \notin E_B$, have penalty factors $p^0_{ij} > 1$. The value of these penalty factors $p^0_{ij}$ may depend on different characteristics of the individiual street segments, representing the perceived safety or convenience.

\begin{figure}[!ht]
    \centering
    \includegraphics{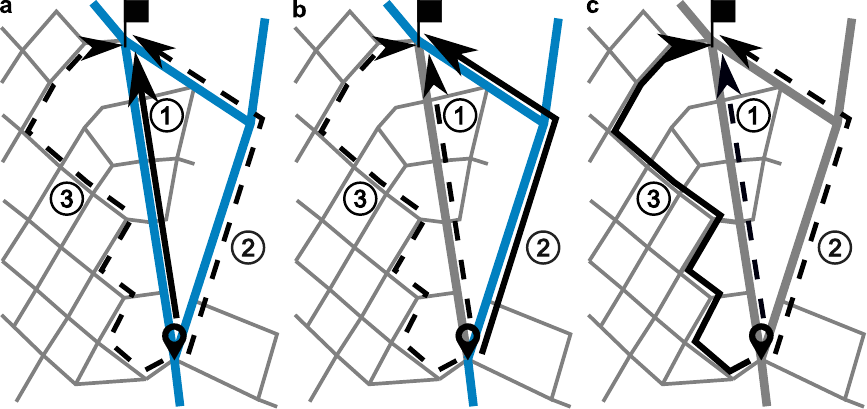}
    \caption{\textbf{Cyclists' route choices trade speed for safety.} 
    Cyclists choose their routes to achieve a short distance and a high safety, potentially avoiding busy streets without a bike path in favor of a longer route with dedicated cycling infrastructure. We model cyclists' route choices as a shortest path problem in a cyclist preference graph [Eq.~\eqref{eq:shortest_path}]: Street segments $e_{ij}$ are assigned an effective distance $l_{ij} = l^\mathrm{street}_{i,j}\,p_{ij}$ combining their physical length $l^\mathrm{street}_{ij}$ and a penalty factor $p_{ij} \ge 1$ summarizing their perceived safety. 
    (a) If all major streets (thick edges) are outfitted with dedicated bike paths (blue), cyclists choose the most direct route (1, solid black arrow) from their origin (pin) to a destination (flag). 
    (b) If only some major streets are equipped with a bike path, cyclists avoid busy roads without a bike path (thick gray) and may prefer a short detour (2). 
    (c) If none of the streets have dedicated bike infrastructure, cyclists balance the distance and safety of their route choices and may prefer long detours via low-traffic residential streets (thin gray, 3) compared to more direct routes with high car traffic (dashed black).
    }
    \label{fig:fig1}
\end{figure}

Adopting this perspective of a cyclist preference graph, we take cyclists to choose their route based on the shortest path $\Pi^*_{i \rightarrow j} = \mathrm{argmin}\left[L_{i \rightarrow j}(\Pi_{i \rightarrow j})\right]$ between their origin and destination, minimizing the perceived trip distance
\begin{equation}
L_{i \rightarrow j}(\Pi_{i \rightarrow j}) = \sum_{e \in \Pi_{i \rightarrow j}} l_e  \label{eq:shortest_path}
\end{equation}
over their potential paths $\Pi_{i \rightarrow j}$. Effectively, cyclists choose the most direct path to keep the physical distance of their trip as small as possible but accept detours to avoid busy streets and use bike paths or low-traffic residential streets as alternative routes (Fig.~\ref{fig:fig1}).

This simplified route choice model enables efficient calculation of route choice decisions, in particular compared to more complex stochastic models \cite{Daganzo1977, Cominetti2010, Storch2020}. To illustrate the concept, we focus here on the effect of street type and take the penalties $p^0_{ij}$ of a street segment to depend on the volume of car traffic on the respective segment, with penalties being higher the larger the car traffic volume (i.e. the lower the perceived safety or convenience), see Methods for details.
In principle, the approach can be extended to include additional factors such as (left) turns and crossings by appropriately modifying the cyclist preference graph (e.g. adding additional edges with a penalty for left turns) or even more complex route choice models or heterogeneous preferences among cyclists \cite{Broach2012_where}. 

\subsection*{Network generation}
We describe the bike path network of a city as a subgraph $G_{B} = (V, E_{B}) \subseteq G$ of the city's street network where each street segment $e_{ij} \in E$ may either be equipped with a bike path, $e_{ij} \in E_B$, or not, $e_{ij} \notin E_B$. Even in this simple binary model, the number of possible bike path networks $G_{B}$ scales exponentially with the number $M$ of edges in the street network since each street segment may be equipped with a bike path or not (thus there are $2^M$ possible subgraphs). Testing all of these networks is impossible for real-world cities in reasonable time. Recent approaches utilize forward network percolation models to construct bike path networks \cite{Natera2020_datadriven, Szell2021_growing} or apply percolation models to a fixed cyclist flow \cite{Olmos2020_data} to find efficient bike path networks. 

\begin{figure}[!ht]
    \centering
    \includegraphics{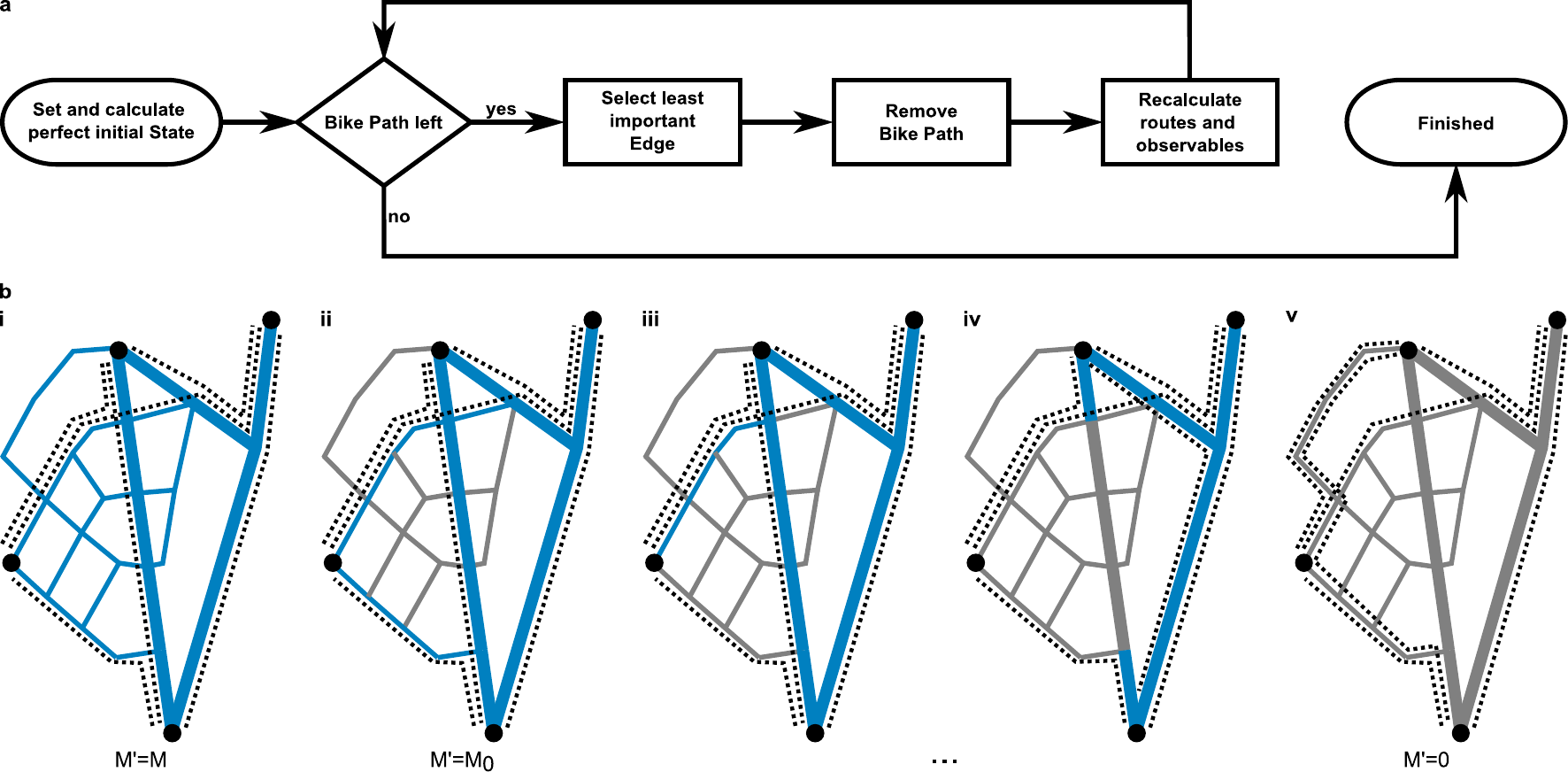}
    \caption{
    \textbf{Constructing a sequence of efficient bike path networks.}  
    (a) Block diagram of the algorithm. 
    (b) Illustration of different bike path networks $G_B(M')$. (i) We start from a full bike path network $G_B(M) = G$, where every street segment of the network $G$ is equipped with a bike path ($M' = M$, blue). (ii) We first remove all unused bike paths that do not affect the cyclists, $p^0_{ij} \, n_{ij} = 0$, leaving us with the smallest subgraph $G_B(M_0)$ that still optimally serves the given demand. (iii, iv) We then one by one remove the least important edges, defined by the smallest product $p^0_{ij} \, n_{ij}$ of penalty factor and number of cyclists using the bike path, recording one network $G_B(M')$ for each number $M'$ of bike paths. (v) The algorithm terminates with an empty bike path network $G_B(0)$ once all bike paths have been removed.
    Upon removal of a bike path from a street segment, we update the routes of all cyclists who used the removed bike path and update the importance rating $p_{ij} \, n_{ij}$ such that we always compute the importance of a bike path with respect to the cycling usage in the current bike path network.
    }
    \label{fig:fig2}
\end{figure}

Here, we employ a complementary approach, following the idea or pruning links from a network, previously employed in network community detection algorithms \cite{Newman2004_finding} and for studying the structure of aviation networks \cite{Verma2016_emergence}. Specifically, we create a sequence $\{G_B(M')\}_{M'}$ of bike path networks where $M' \in \{0,1,\dots,M\}$ street segments are outfitted with a bike path (see Fig.~\ref{fig:fig2}): We start from an optimal bike path network $G_B(M) = G$ where every street segment is equipped with a bike path, $E_B(M) = E$, such that there is no penalty for any street segment, $p_{ij} = p^B_{ij} = 1$ for all edges. We then compute the route choice decisions of the cyclists in their preference graph $G$ as described above based on their demand distribution $n_{i \rightarrow j}$. Here, $n_{i \rightarrow j}$ denotes the number of cyclists traveling from node $i$ to $j$. To construct the family of bike path networks, we one by one remove the least important bike path $e^*_{ij}(M')$ from the network, $E_B(M' - 1) = E_B(M') \setminus \{e^*_{ij}(M')\}$, adjusting the penalty of that street segment from $p_{ij} = p^B_{ij} = 1$ to $p_{ij} = p^0_{ij} > 1$ in the cyclist preference graph $G$. We quantify the importance of a bike path $e_{ij} \in E_B(M')$ in the current state of the bike path network (with $M'$ remaining bike paths) as the product $p^0_{ij} \, n_{ij}(M')$ of penalty $p^0_{ij}$ if the street had no bike path and the number of current users of that street segment $n_{ij}(M')$. The product represents the graph-theoretical weighted betweenness centrality of the edge in the cyclist preference graph. This approach minimizes the negative impact of each removed bike path on the perceived distance of the cyclists in the current bike path network. After each change of the cyclist preference graph $G$, we update the route choice decisions of the cyclists, ensuring that the algorithm continually adapts to the cycling demand given the currently available set of bike paths $E_B(M')$. For example, it keeps bike paths that may not be important in the perfect network if cyclists start to use them more heavily as other bike paths are removed. 
The process terminates with an empty bike path network $G_B(0)=(V,\emptyset)$ once all bike paths have been removed.

In contrast to iteratively adding bike paths to an initially empty graph and building on the suboptimal cycling routes in networks with few bike paths, this procedure creates bike path networks adjusted to the ideal cycling conditions. For example, instead of constructing bike paths in residential areas with a high number of cyclists because large streets are not yet outfitted with bike paths, it keeps large streets equipped with bike paths in the first place and would only remove bike paths from segments of large streets if they are not important for a direct trip for many cyclists. 

Inputs to our algorithm are (i) the street network $G^\mathrm{street}$, (ii) the penalty factors $p^0_{ij}$ for each street segment not equipped with a bike path, (iii) the demand distribution $n_{i\rightarrow j}$, and (iv) the cyclists' route choice model. These parameters may either be as-is empirical values or planned or desired ideal values (e.g. describing the desired or predicted demand of cycling usage in a city). The latter application might be particularly relevant for planning bike path network extensions if urban quarters develop or are repurposed.

For the computational runtime of the network generation, see Appendix.

\clearpage

\subsection*{Application}

\begin{figure}[ht]
    \centering
    \includegraphics{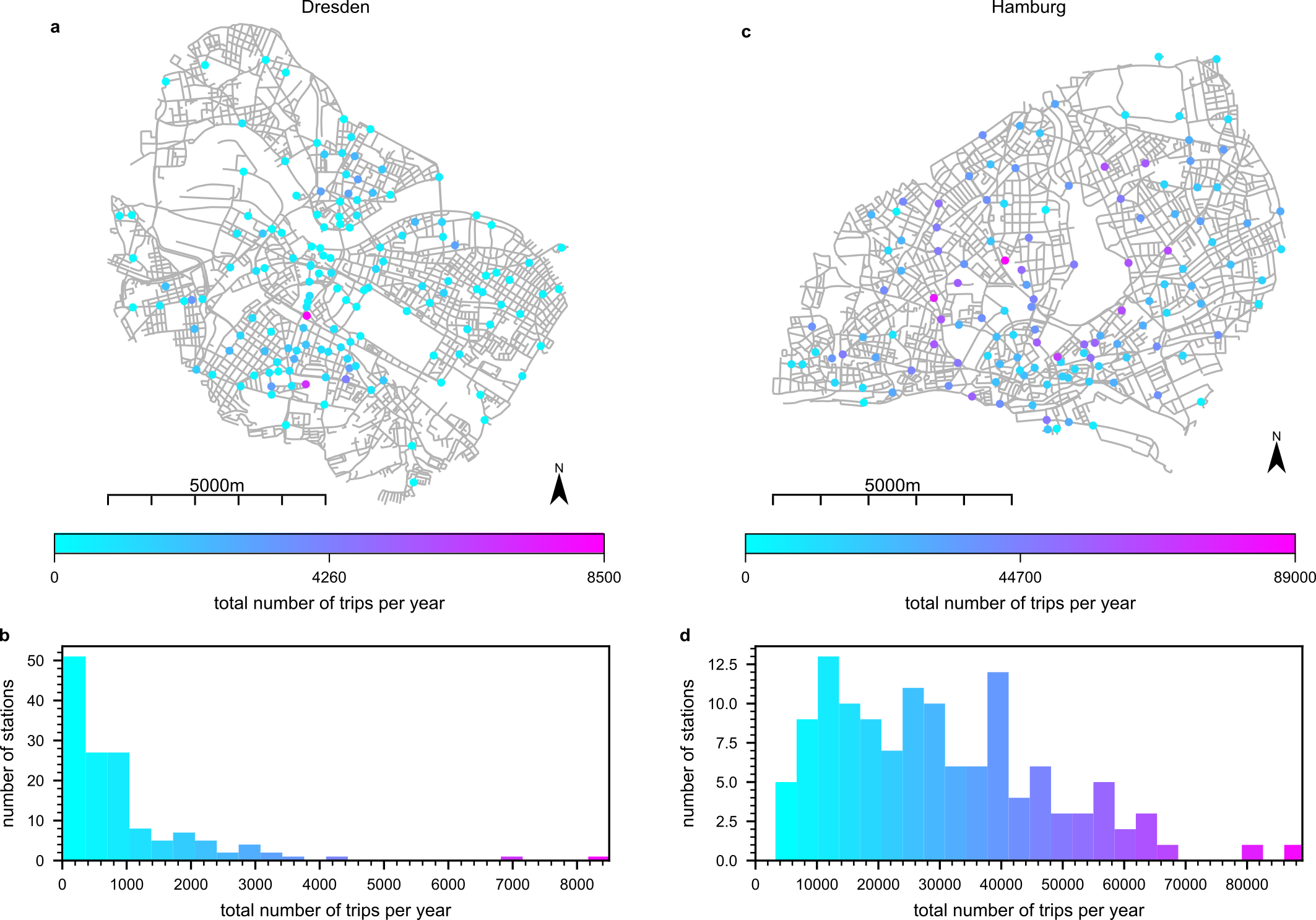}
    \caption{
    \textbf{Street networks and bike sharing demand in Dresden and Hamburg.}
    (a) Street network (gray lines), bike sharing station locations, and station activity (circles) in Dresden between November 2017 to March 2020 \cite{Nextbike}. 
    (b) Distribution of station usage, measuring the combined number of in- and out-going trips per station, in Dresden. Bike sharing usage is strongly heterogeneous, dominated by two heavily used stations (pink) along the north-south axis between the central train station (center) and the university campus (south). Station density reflects this usage pattern and is highest in the central city (north/center) and near the university campus (south). 
    (c) Street network (gray lines), bike sharing station locations, and station activity (circles) in Hamburg between January 2014 to May 2017 \cite{DBData}. 
    (d) In Hamburg, the station activity distribution is homogeneous across a broad spectrum of total number of trips. This homogeneous usage is also reflected in a more homogeneous distribution of bike sharing stations, which is slightly denser only in the inner city (south).
    }
    \label{fig:fig3}
\end{figure}

We test the proposed algorithm with data from two German cities, Dresden and Hamburg (see Fig.~\ref{fig:fig3}). As input data we take the street network of both cities from OpenStreetMap (OSM) \cite{OSM}, including the classification of the streets as a proxy for their expected traffic load, and data from local bike sharing services to model the cycling demand. We fix the penalty factors compared to physically protected bike path infrastructure based on the street type classification decoded in OSM, assuming street types to be a proxy for the traffic load on a street. See Methods for a detailed description of the data. 

The two cities are representatives of two archetypes of local demand constellations: spatially homogeneous all-to-all and confined few-to-few demand. Bike sharing usage patterns in Hamburg indicate a local demand structure referring to the first archetype (see Fig.~\ref{fig:fig3}c,d). Corresponding data for Dresden hint at the latter archetype, reflected by strong dominance of trips between university and main train station (see Fig.~\ref{fig:fig3}a,b).

\clearpage

\subsubsection*{Algorithmic generation of bike path networks}

We generate families of bike path networks $\{G_B(M')\}_{M'}$ for both cities. As a comparison we chose a network where all primary and secondary (P+S) street segments (as per their classification in OSM) are equipped with bike paths. We compare this network to our generated network with the same total length $\Lambda(M') = \sum_{e \in E_B(M')} l^\mathrm{street}_{e}$ of bike paths such that $\Lambda(M') = \Lambda_\mathrm{P+S}$. Taking the cost for installation and maintenance of the bike path network proportional to its length, we thus compare networks for the same budget. Due to some antiparallel one-way streets, our algorithm may place slightly more bike paths than effectively exist in the primary and secondary network. Since we assume bidirectional paths, our algorithm may equip only one of the antiparallel streets with a bike path instead of both as in the P+S network.

\begin{figure}[ht]
    \centering
    \includegraphics{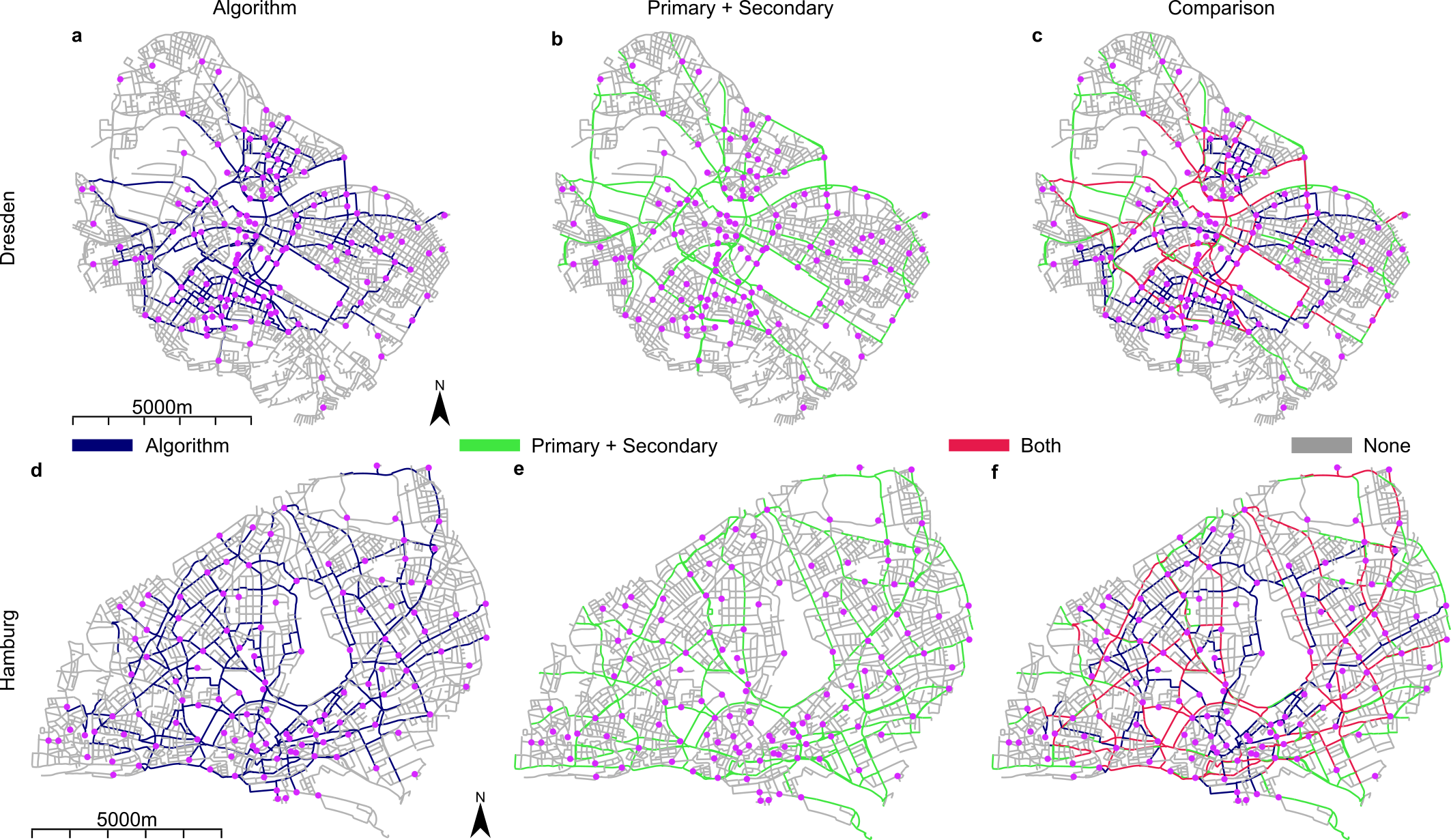}
    \caption{
    \textbf{Demand-efficient bike path networks.} 
    (a-c) Dresden, (d-f) Hamburg. Panels a and d show the networks generated by the algorithm with a total length $\Lambda(M') = \Lambda_{\mathrm{P+S}}$. Panels b and e show networks with the same total length if only the primary and secondary streets (as per their OSM classification) are equipped with bike paths. Panels c and f show a comparison between both networks. 
    The network generated by the proposed algorithm largely coincides with the primary-secondary-network (orange edges in panel c, f). However, it more accurately reflects the input demand structure, keeping also highly used tertiary or residential streets equipped with bike paths and exhibiting a higher density of bike paths in high-demand areas. This is particularly visible by the concentration of bike paths along the central areas with high station density in Dresden. The differences are smaller in Hamburg due to the more homogeneous demand structure.
    }
    \label{fig:fig4}
\end{figure}

Figure~\ref{fig:fig4} illustrates both types of networks for Dresden and Hamburg. The network generated by our algorithm largely coincides with the primary and secondary roads due to the high penalty if bike paths were removed. However, we observe strong differences in the density of the bike path coverage. Especially in Dresden, the resulting bike path network is much denser along the central north-south axis of high station density and bike-sharing usage, indicating that our algorithm correctly adapts the network to the input demand conditions (compare Fig.~\ref{fig:fig4}c). The differences for Hamburg are smaller due to the comparatively homogeneous demand across the city, though our algorithm introduces bike path shortcuts through residential areas in cases of high demand or to connect stations to the bike path network.

\clearpage
To quantitatively compare the bike path families for both cities, we normalize the length of bike paths $\lambda = \Lambda(M') / \Lambda(M_0)$ with respect to the length $\Lambda(M_0)$ after removing all unused bike paths (compare Fig.~\ref{fig:fig2}b). We define the total perceived distance of all trips in the cyclist preference graph as 
\begin{equation}
    \mathcal{L}(\lambda) = \sum_{i,j \in V} n_{i \rightarrow j} \, L_{i \rightarrow j}(\Pi^*_{i \rightarrow j}(\lambda),\lambda)
\end{equation} 
with $L_{i \rightarrow j}(\Pi,\lambda) = \sum_{e \in \Pi} l_{e}(\lambda)$ [compare Eq.~\eqref{eq:shortest_path}].  Here $l_{e}(\lambda)$ denotes the effective length of the street segment $e$ in the cyclist preference graph $G$ given a set of bike paths $E_B(M')$ with normalized length $\lambda$, i.e. including penalties only for those streets where we have removed the bike path. $\Pi^*_{i \rightarrow j}(\lambda)$ denotes the shortest path in this cyclist preference graph and thus the route chosen by cyclists going from $i$ to $j$. To compare the total perceived distance across both cities, we measure the overall performance $b(\lambda)$ of the resulting network as the bikeability
\begin{equation}
    b(\lambda) = \frac{\mathcal{L}(0) - \mathcal{L}(\lambda)}{\mathcal{L}(0) - \mathcal{L}(1)}, \label{eq:bikeability}
\end{equation}
where we again normalize the absolute values to the best case ($\lambda=1$) and worst case ($\lambda=0$) scenarios. $b(0) = 0$ describes the network with no bike paths and $b(1) = 1$ is the optimal network with bike paths along all shortest paths. 
We remark that our definition of bikeability differs from previous measures of bikeability \cite{Kellstedt2021_bikeability} in that it quantifies the efficiency of the bike path network with respect to a specific demand distribution. 
The total difference between the physical trip length of cyclists and the direct shortest paths is only of the order of $10\%$, consistent with the empirical observations of cyclists' route choice behavior \cite{Menghini2010_routechoice, Broach2012_where}.

\begin{figure}[ht]
    \centering
    \includegraphics{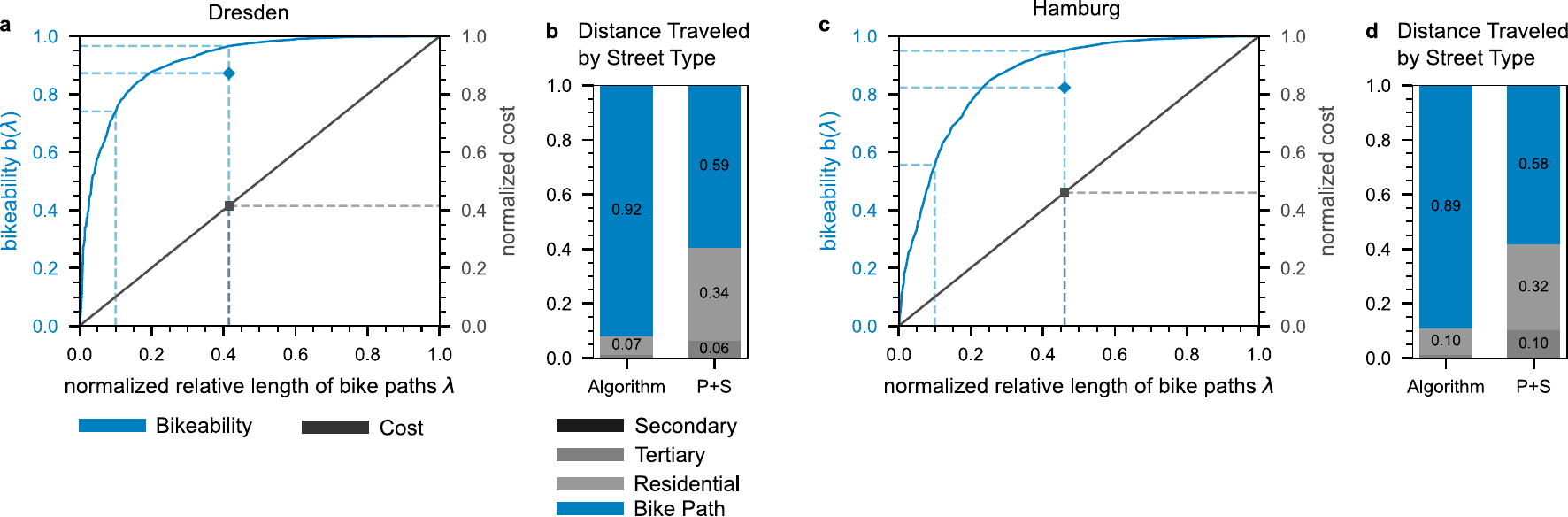}
    \caption{
    \textbf{Demand-driven design of bike path networks improves bikeability.} 
    Resulting bikeability for (a, b)  Dresden and (c, d) Hamburg as a function of the fraction of street segments equipped with bike paths.
    In both Dresden (a) and Hamburg (c), the bikeability $b(\lambda)$, quantified via the normalized effective distance of routes in the cyclist preference graph [Eq.~\eqref{eq:bikeability}], increases quickly as the length of bike paths in the network grows. In contrast, the relative cost of the network, taken to be proportional to the total length of bike paths, increases linearly. Even a low density of bike paths $\lambda > 0.10$ increases the bikeability to $b(\lambda) > 0.50$, with a larger effect of a small fraction of bike paths in Dresden. Equipping only the largest streets (primary and secondary, blue circle in panels a and c, $b_\mathrm{P+S} = 0.87$ for Dresden and $b_\mathrm{P+S} = 0.82$ for Hamburg) already achieves a substantial bikeability due to avoiding high penalties of these streets. Our demand-efficient bike path networks close more than $70\%$ of the remaining gap to the best possible bikeability (if all streets are equipped by bike paths) at the same cost by adapting the bike path network to the input demand, achieving $b(\lambda_{\mathrm{P+S}}) = 0.97$ for Dresden and $b(\lambda_{\mathrm{P+S}}) = 0.95$ for Hamburg.
    (b, d) Fraction of total distance traveled on streets with and without bike path in networks with bike paths with a relative length $\lambda_{\mathrm{P+S}}$ and the primary-secondary (P+S) comparison network. Adapting the network to the demand structure achieves over $89\%$ of the distance traveled on bike paths due to equipping also smaller streets used by many cyclists with dedicated bike infrastructure (compare Fig.~\ref{fig:fig4}). For our algorithm, a negligible fraction of the total distance is cycled on tertiary and secondary roads without a bike path (not visible in the bar chart).
    }
    \label{fig:fig5}
\end{figure}

Fig.~\ref{fig:fig5} illustrates the bikeability across the generated sequence of bike path networks. Interestingly, already a small fraction of bike paths with a small relative length $\lambda > 0.1$ is sufficient to achieve more than $50\%$ of the maximal bikeability in both cities. The larger area under the bikeability curve for Dresden compared to Hamburg is consistent with the differences in the demand structure between the two cities: we achieve a faster improvement in Dresden due to the more concentrated demand distribution whereas we have to cover most of the city in Hamburg due to the more homogeneous demand. For a brief overview of the bikeability for 12 additional cities, see  Fig.~\ref{fig_app:cities}.

A comparison with the bikeability of the primary-secondary bike path network with the same relative length $\lambda_{\mathrm{P+S}}$ of bike paths highlights the better adaptation to the demand structure in our algorithm. The bikeability is already high when all large roads are equipped with bike paths, about $0.87$ for Dresden and $0.82$ for Hamburg. Yet, our algorithm manages to further increase this value to about $0.97$ for Dresden and $0.95$ for Hamburg (capturing more than $70\%$ of the remaining potential of an optimal network $b(1) = 1$). Moreover, by adjusting the network to the route choice behavior, cyclists keep to streets equipped with bike paths for more than $89\%$ of their total trip distance, compared to only about $60\%$ in the primary-secondary network (see Fig.~\ref{fig:fig5}b and d). A negligable but non-zero fraction of the distance is cycled on tertiary and secondary streets without a bike path. For a comparison to static percolation approaches, see Appendix.

\subsubsection*{Impact of demand structure}
\begin{figure}[ht]
    \centering
    \includegraphics{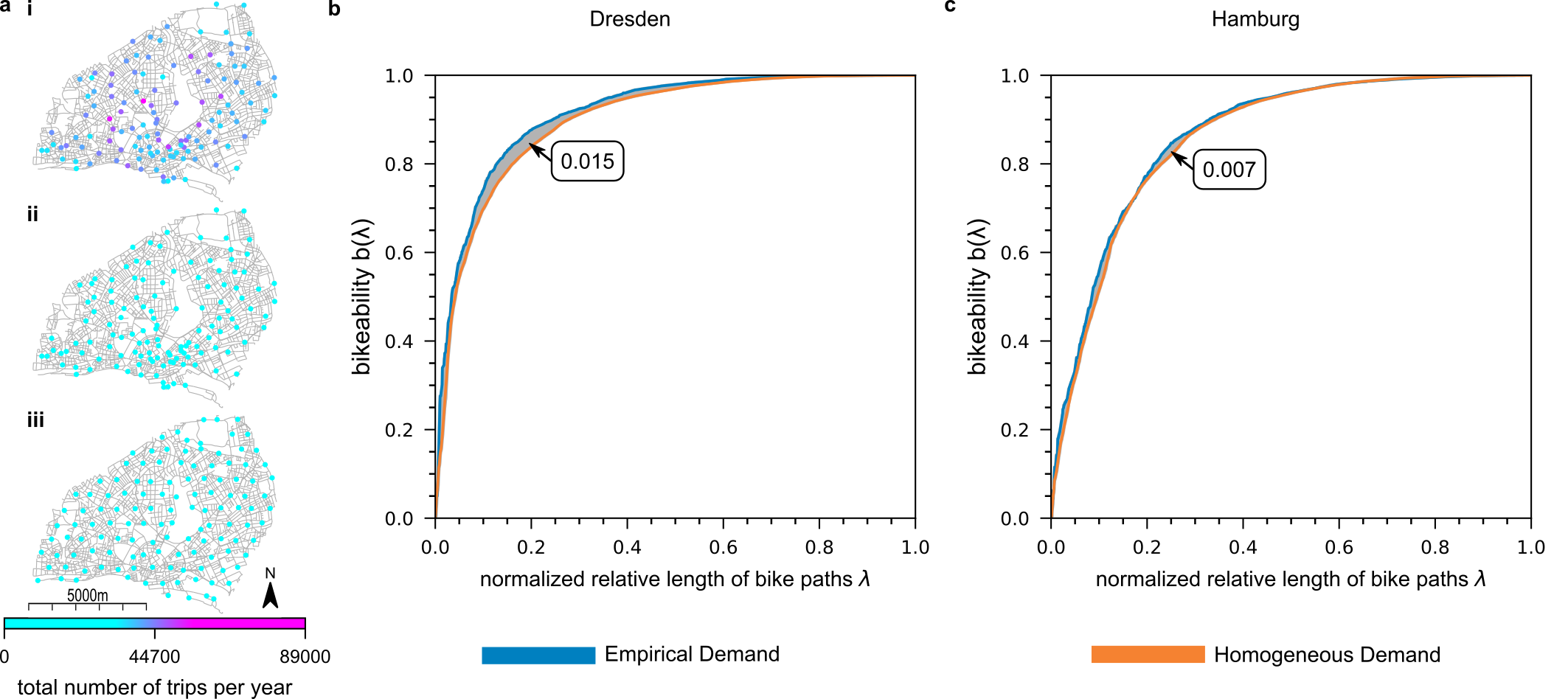}
    \caption{
    \textbf{Bikeability differences quantify the importance of demand structures.}  
    (a) Illustration of the demand homogenization for Hamburg. Starting from the empirical demand (i), we create uniform demand between all stations (ii) and then distribute the stations approximately equidistantly across the city (iii). 
    (b, c) Comparison of the bikeability from the empirical and the homogeneous demand data for Dresden and Hamburg, respectively. The comparatively large area between the bikeability curves of empirical and homogeneous demand in Dresden [$\beta_\mathrm{hom} \approx 0.015$, compare Eq.~\eqref{eq:area}] suggests a large effect of the more localized demand distribution, making it easier to create an efficient bike path network (compare Fig.~\ref{fig:fig4}). For Hamburg, we find a two times smaller change ($\beta_\mathrm{hom} \approx 0.007$) confirming our earlier observations that the empirical demand distribution is already largely homogeneous across the city. 
    }
    \label{fig:fig6}
\end{figure}

We attribute the difference between the two cities in the above analysis to the structure of the bike-sharing demand distributions. To quantify the impact of the demand structure on our bike path network families and their bikeability curves, we compare the previous results to synthetic bike path networks for homogenized demand. We create these homogeneous demand settings by first distributing demand equally between all stations and then distributing the stations as equidistantly as possible in the street network (see Fig.~\ref{fig:fig6}a and Methods).

Comparing the bikeability curves $b(\lambda)$ and $b_\mathrm{hom}(\lambda)$ in the empirical and the homogeneous demand settings, respectively, we find a comparatively large difference for Dresden and a much smaller difference for Hamburg (Fig.~\ref{fig:fig6}b,c). We quantify these differences by the area 

\begin{equation}
    \beta_\mathrm{hom} = \int_0^{1} \left[b(\lambda) \, - b_\mathrm{hom}(\lambda)\right] \, \mathrm{d} \lambda \,, \label{eq:area}
\end{equation}

between the two bikeability curves, describing the impact of the patterns and the structure in the station and demand distribution on the bikeability ($\beta_\mathrm{hom} \approx 0.015$ for Dresden, $\beta_\mathrm{hom} \approx 0.007$ for Hamburg, compare Fig.~\ref{fig:fig6}). This confirms our above analysis that the heterogeneous, centralized demand and station distribution in Dresden enhances the bikeability as fewer streets have to be equipped with bike paths to cover a large fraction of the total demand, while there is a significantly smaller effect in Hamburg. A similar approach may be used to quantitatively compare the impact different street networks, different types of cycling demand, or different desired demand distributions may have on the resulting bike path networks.

\vspace{-0.5cm}
\section*{Discussion}
Promoting cycling as a way to improve the sustainability of urban mobility is a complex problem. The attractiveness of cycling depends on the available infrastructure and safety of the trips, alternative mobility options, other mode choice decisions influencing the amount of car traffic, as well as aspects like topography and weather \cite{Munoz2013_topo, Gebhart2014_weather, Saneinejad2012_weather}.

Existing approaches to study bike path networks focus on different properties and rely on various types of input data. Using only existing bike paths as input, \cite{Natera2020_datadriven} focuses on the structural connectivity of these networks by linking the fragmented components without accounting for demand. Taking as input a qualitative demand distribution in the form of (artificial) points of interest, \cite{Szell2021_growing} considers network connectivity and physical shortest path routing to grow bike path networks without accounting for different types of streets or the route choices of cyclists. Building on empirical demand data and a detailed but static route choice model in an existing network, \cite{Olmos2020_data} considers only the number of cyclists on each street segment (compare Fig.~\ref{fig_app:dyn_stat}), without accounting for route choice changes as the network grows or for different penalty weighting across different street types. 

We have introduced an adaptive percolation framework to generate families of bike path networks that includes cycling demand as well as adaptive route choices of cyclists as the network changes by using a simplified, penalty based route choice model \cite{Broach2012_where, Menghini2010_routechoice}. Compared to more abstract percolation models, our approach trades computational speed for the explicit inclusion of cyclist demand. Compared to more detailed models, we trade accuracy of the route choice model for the ability to adaptively adjust route choices as the network evolves. The resulting bike path networks highlight that most of the potential effect of bike paths on cycling quality may be achieved already with a small investment in cycling infrastructure. Moreover, the framework enables us to quantify the impact of the network topology and demand distribution by providing a direct way to compare the resulting networks in different settings \cite{Buehler2011_impact_bikepaths, Munoz2013_topo}. 

Similar network pruning or inverse percolation techniques, as for cascading failure models \cite{Carmona2020_cracking, Schafer2018_Cascading}, have previously been applied to understand the structure of transportation networks and their robustness \cite{Verma2016_emergence, Barthelemy2011_spatial}. The approach may thus find applications in designing and analyzing infrastructure and transportation networks also beyond cycling. Combining the suggested approach with additional optimization steps, for example using successive addition and removal of individual bike paths or simulated annealing techniques \cite{Kirkpatrick671_Annealing, Kirkegaard2020_annealing, Gastner2006_optimal} to explore the vicinity of networks constructed by our greedy approach, may further improve the quality of the resulting networks.

The proposed framework relies on several types of input, all with potential limitations in terms of data quality, modelling accuracy, and interpretability of the results. However, the framework is highly adaptive and can easily be extended to overcome these challenges provided more detailed input data is available.

Firstly, the quality and type of input data is critical for the resulting networks and their interpretation. Available street network data is mostly of high quality (see Methods) and even existing bicycle infrastructure may be included in the framework by preventing the removal of bike paths from specific street segments in the network. However, the resulting bike path networks have to be interpreted in the context of the demand input (compare our results for Dresden). For example, the empirical bike sharing demand used to illustrate the framework may be strongly influenced by the currently (non-)existing infrastructure and the type of users of the service, thus skewing the generated networks to further improve already efficient parts of the network. At the same time, the framework is not limited to empirical demand data. Constructing efficient networks for desired and predicted demand may help guide extensions of bicycle infrastructure networks by suggesting efficient network structures \cite{Creutzig2016, Olmos2020_data}. Combining full mobility demand with a suitable mode choice model may even enable us to capture effects of induced additional demand when sufficient infrastructure is provided \cite{OECD2021, Storch2020}.

Secondly, the framework relies on a simplified route choice model to enable fast computation by mapping the route choice of cyclists to an effective shortest path problem. We illustrated the framework with an effective network capturing only the effect of car traffic volume on cyclists route choice. A more detailed definition of the penalties, including additional deterrents such as slopes and (left) turns, may improve the accuracy of the route choice model. Additionally, explicit safety considerations such as accident risk may be included in the penalties as proposed in \cite{Folco2022_safety}, enabling more accurate quantification of the actual safety in addition to the perceived safety. A more accurate representation of cyclist route choice behavior and heterogeneous preferences among cyclists, as assumed in common probabilistic route choice models, may be indirectly possible by considering multiple types of cyclists and creating a cyclist preference graph with appropriate penalties for each user type. 

All of these potential extensions naturally require more details in the input data, such as information on traffic signals, street quality, or expected driving behavior for car traffic across the street network. Recent contributions to the data-driven analysis and planning of cycling infrastructure and route choice as well as data collection methods are essential to establish a foundation of input data and ensure reliable results and predictions \cite{Olmos2020_data, Natera2020_datadriven, Szell2021_growing}. While eventually first-hand on-site experience and detailed case-by-case modelling must determine the sensibility and feasibility of the suggested networks, with sufficiently accurate input data our framework may provide scenarios for efficient bicycle infrastructure networks, helping to guide planned extensions of existing infrastructure networks \cite{Creutzig2016, Olmos2020_data}. Our framework may thus complement current urban planning approaches \cite{RikdeGroot2016_designmanual} by (i) helping to develop a more detailed understanding of the theoretical properties of efficient bike network structures across cities and (ii) offering a baseline of an efficient network to develop a more detailed long-term strategy to expand bicycle infrastructure.

Overall, the framework presented in this article may enable quantitative analysis of bike path networks with a large range of tools from network science by providing a way to quickly generate and compare families of efficient bike path networks in different settings and under different conditions.

%\newpage 

\section*{Methods}

\subsection*{Street networks}
We download physical street networks for Hamburg and Dresden from OpenStreetMap \cite{OSM, OSMnx}.
Although OSM data is crowdsourced, it is of high quality in developed countries, especially in Western Europe \cite{Graser2014_osmquality}. For other regions of the world, OSM is sometimes the only feasible source of data \cite{Quinn2019_osmquality}.

For both cities, we restrict ourselves to the area covered by the local bike-sharing schemes. We exclude city peripheries that are either sparsely populated with low density of bike sharing stations or a long distance from the city center. This results in a reasonably bikeable area of approximately 65 km$^2$ (Dresden) and 49 km$^2$ for Hamburg (Fig.~\ref{fig:fig3}).

We simplify this raw street networks by merging nodes within a radius of less than 35 m (e.g. simplifying the detailed structure of intersections), and placing the resulting aggregated node at the centroid of their former position \cite{OSMnx}. Finally, we discard bike inaccessible roads based on the OSM street classification hierarchy \cite{OSMwiki}, in particular edges labeled as 'motorway', 'motorway\_link', 'trunk', 'trunk\_link', describing highways or high-speed motorways where cycling is not possible.

Based on the remaining nodes and edges (see Fig.~\ref{fig:fig3}) we create the physical street network $G^\mathrm{street}$ by assigning each street segment a length $l^\mathrm{street}_{ij}$ corresponding to its physical length in the OSM data.

\subsection{Street size classification} 
To fix the penalty factors in these networks without detailed information on traffic volume, we rely on the street type classification decoded in OSM as a proxy for the traffic load. Within the OSM-category of car-accessible streets, we use OSM's classification hierarchy into 'primary', 'secondary', 'tertiary', or 'residential' streets, replacing link street types with their corresponding normal street types, e.g. 'primary\_link' with 'primary. Street segments not labeled with one of the four aforementioned classifications, will be assigned the 'residential' status, as most other OSM classifications are reserved for small streets, e.g. 'living\_street'. In case of ambiguity about the street segment length or street type, we always choose the first value in the list of the OSM data. 

We assume that the expected traffic load and thus the corresponding penalty factor $p^0_{ij}$ monotonically increases from residential to primary roads (Tab.~\ref{tab:street-penalty}). Based on this input, we construct the cyclist preference graph $G$ with the perceived edge lengths $l_{ij} = p_{ij}\,l^\mathrm{street}_{ij}$ [Eq.~\eqref{eq:weighted_edge_distance}].

The penalty factors $p_{ij}^0$ quantify the trade-off between the distance a cyclist is willing to ride along a street with a bike path in order to avoid a specific street segment without bike path. To illustrate our framework, we take penalty values representing the perceived distance compared to physically protected bike paths as they are the safest option for cyclists. However, the penalties can be adapted to represent other types of bike path implementations. We employ penalty factors as estimated in Ref. \cite{Broach2012_where}; the authors associated different penalties with different properties of the route and street segment by contrasting actually chosen bike routes with possible alternatives in a logit route choice model. We take the penalty factors representing accepted detours for streets with more than $30$k vehicles per day, $20$k to $30$k vehicles per day and $10$k to $20$k vehicles per day, respectively. Due to lack of detailed information on traffic volume, we directly match these penalties to the OSM street types (see results in  Tab.~\ref{tab:street-penalty}). For 'residential' street segments, we add a smaller penalty factor based on the other three values. These values serve as illustrative examples of the influence of car traffic and may vary depending on the location, trip purpose, bicycle infrastructure, or for individual cyclists.

\begin{table}[ht]
\centering
\begin{tabular}{ll}
street type & penalty $p^0_{ij}$ \\
primary & 7.0 \\
secondary & 2.4 \\
tertiary & 1.4 \\
residential & 1.1
\end{tabular}
\caption{\textbf{High traffic volume streets are perceived longer by cyclists}. Depending on the OSM street type classification hierarchy for street size, we determine cyclists' perceived street segment length loosely based on empirical length penalty factors \cite{Broach2012_where}.
} \label{tab:street-penalty}
\end{table}

\subsection*{Bike sharing demand data} 
We take demand data from local bike sharing services in Dresden and Hamburg as demand input for the algorithm (Fig.~\ref{fig:fig3}). 
For both cities, we map the bike sharing stations to their respective nearest node in the street network and obtain the origin-destination resolved demand statistics by counting the overall number of trips $n_{i \rightarrow j}$ made throughout the entire observation period per pair $(i,j)$ of stations. To quantify the total usage of each station, we compute the sum of bike rentals and returns at the station (compare Fig.~\ref{fig:fig3}b,d). The two datasets represent two archetypes of local demand constellations: confined few-to-few demand in Dresden (Fig.~\ref{fig:fig3}a,b) and spatially homogeneous all-to-all demand in Hamburg (Fig.~\ref{fig:fig3}c,d). 

\subsubsection*{Dresden} 
Dresden's local bike-sharing scheme operated on a combined station-based and free-floating mode during the observation period. While bikes could be rented or returned on an as-needed basis within a pre-defined area in the inner city center, they needed to be rented from and returned to one of 159 stations outside the free-floating zone \cite{Nextbike}.

To estimate the local bike demand we use a proprietary dataset of approximately 440,000 bike-sharing trip records conducted by students of the local universities between November 2017 and March 2020, except February and September 2018, accounting for about 80\% of all trips of the service \cite{Nextbike}. The dataset contains, among others, information on trip origin and destination if the trip started or ended at a station as well as pickup and drop-off timestamps. The dataset does not contain positional information on trips conducted in free-floating mode.

To be able to fix the demand distribution to nodes in the street network, we exclude trips for which no origin or destination information is available (for example trips starting or ending in the free-floating zones). Furthermore, we exclude 9 stations distant from the city center and thus not included in the core polygon illustrated in Fig.~\ref{fig:fig3}a, leaving approximately 163,000 trip records for the demand analysis. The remaining 150 stations are mapped to 142 nodes of the street network $G$ ($8$ stations are mapped to the same node as another station, e.g. two stations on two sides of a large street crossing).

\subsubsection*{Hamburg} 
In 2017, Hamburg's station-based bike-sharing scheme operated 206 stations of which 129 were distributed in the core city (see Fig.~\ref{fig:fig3}b). Between January 2014 and May 2017 the service facilitated approximately 8.6 million rides for which detailed trip information is publicly available \cite{DBData}. The data contains, among others, information on trip origin and destination station, pickup and drop-off timestamps, as well as user or bike-related information. 

We again exclude trips where no origin or destination information is available as well as trips which start and end at the same station, leaving about 6.4 million trips in our region of interest (approximately 74\% of all trips). After mapping the $129$ stations to the street network $G$, we keep $127$ unique locations ($2$ stations are mapped to the same node as another station).

\subsection*{Homogenized Demand} 
To quantify the impact of the demand distribution on the bikeability, we compare our results for the empirical demand distribution with results for homogenized demand distributions. We generate these randomized comparisons in three steps: (i) create approximately homogeneous demand and station distribution, (ii) generate bike path ensembles for ten realizations of the demand and station distribution, (iii) average the bikeability results from the different realizations.

We create these homogeneous demand settings by first distributing demand equally between all stations, setting $n^\mathrm{(\mathrm{hom})}_{i\to j}=1$ for all bike sharing station pairs $i,j$ with $i \neq j$. Second, we distribute the stations as equidistantly as possible in the street network. To achieve this station distribution, we create a triangular lattice in the polygon of the physical street network with a slightly higher total number of lattice points. We then map each lattice point to the closest node in the underlying street network $G$ and delete excess points starting with those lattice points whose position in the triangular lattice is furthest from its corresponding node in $G$ until we are left with the same number of nodes as station in our original data (Fig.~\ref{fig:fig6}a).

We create bike path networks as described above, compute the resulting bikeability and other measures, and average them over ten realizations of random homogeneous station distributions.

\section*{Data availability}

The bike-sharing trip records used to estimate the cycling demand in Hamburg are publicly available from Ref.~\cite{DBData}. The specific data used for Hamburg in this paper can be found in the GitHub repository \cite{GitHub_repo} with the source code. The bike-sharing trip records for Dresden are a proprietary asset of the Studierendenrat of the Technische Universit\"at Dresden and nextbike.

\section*{Code availability}

The simulation code for our algorithm and a guide to reproducing the results is available through GitHub (\url{https://www.doi.org/10.5281/zenodo.6602117}) \cite{GitHub_repo} under a AGPL-3.0 license.

\section*{Acknowledgements}
We thank Stefan Huber, Sven Li{\ss}ner and Regine Gerike for stimulating discussions. 
C.S. acknowledges support from the German Federal Environmental Foundation (Deutsche Bundesstiftung Umwelt, DBU). 
D.S. acknowledges support from the Studienstiftung des Deutschen Volkes. M.T. acknowledges support from the German Research Foundation (Deutsche Forschungsgemeinschaft, DFG) through the Center for Advancing Electronics Dresden (cfaed). The project was partially funded by the Deutsche Forschungsgemeinschaft (DFG, German Research Foundation) – project number 493613373 to M.S. .

\section*{Author contribution} D.S. and M.S. initiated the research supported by M.T. All authors conceived and planned the research. C.S. collected and analyzed the empirical data supported by D.S. C.S. and M.S. designed the algorithm. C.S. wrote the code, performed and analyzed the simulations supported by M.S. All authors contributed to interpreting the results and writing the manuscript.

\section*{Competing interest}
The authors declare no competing interests.

\newpage
\section*{Appendix}
\appendix

\renewcommand{\thefigure}{A\arabic{figure}}
\setcounter{figure}{0}

\subsection*{Demonstration for additional cities} \label{sec:app_1}

\begin{figure}[h]
    \centering
    \includegraphics{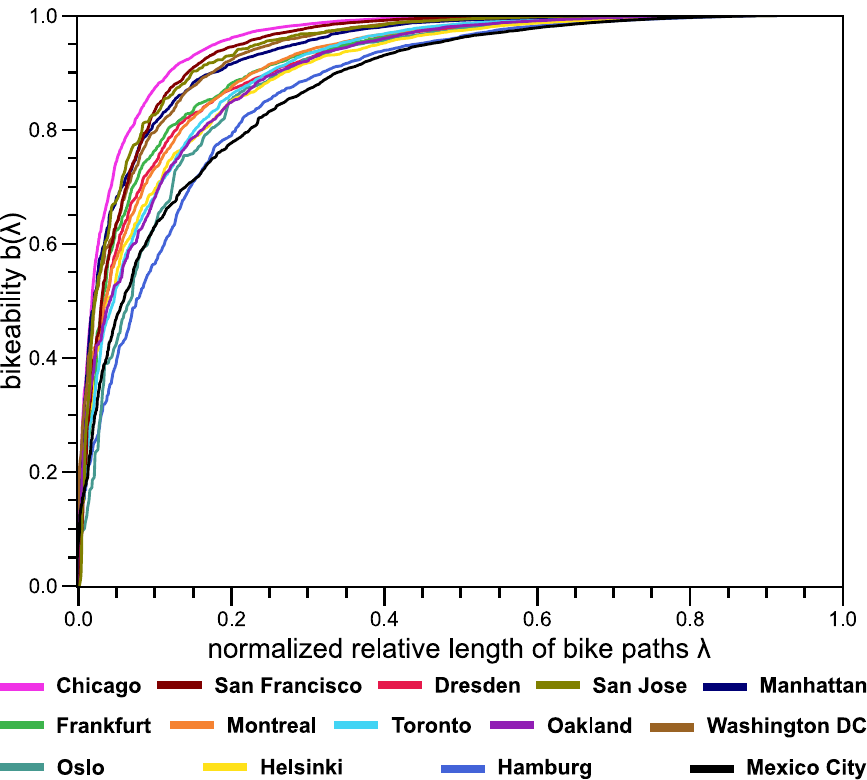}
    \caption{\textbf{Fast improvements of bikeability across cities.} The bikeability $b(\lambda)$ shows a qualitatively similar behavior across 14 different cities. In all cities, the dynamic percolation framework creates large improvements in bikeability with relatively few bike paths (compare Fig.~5 in the main manuscript). The differences in bikeability of the cities likely result from a variety of factors, including the demand distribution or size and topology of the network, required to accurately contextualize the results for each city (compare the detailed discussion for Hamburg and Dresden in the main manuscript).
    }
    \label{fig_app:cities}
\end{figure}

Applying our algorithm to 12 additional cities (see \cite{ChicagoData, BayAreaData, NYCData, DBData, MontrealData, TorontoParkingAuthority, WashingtonData, OsloBysykkel, HelsinkiData, MexicoCityData} for the underlying data), we find qualitatively the same results for all cities. Bikeability quickly increases with a small fraction of bike paths for all cities (see Fig.~\ref{fig_app:cities}). Differences between the bikeability curves of the cities stem from the different demand distributions and number of stations in the available bike-sharing data used as demand input for the network generation. Additionally, the size of the network and potentially inconsistent classification of the streets in Open Street Map across different cities may also affect the algorithm. A more detailed analysis of all these factors would be required to accurately contextualize the results for each city.

We especially note that directly comparing the bikeability of different cities is not straightforward. Since the bikeability measures the normalized improvement from the worst network state (without any bike paths) to the optimal network state, high bikeability may mean different things. Bikeability may be high, because the network is very bikeable, e.g. due to very concentrated demand requiring few bike paths, such as along the North-South axis of Manhattan. However, it may also mean a very non-bikeable network in the absence of a bike path network and thus a proportionally higher increase as bike paths are added (e.g. Chicago). An accurate interpretation of the bikeability measure thus requires the context of the street network and the demand.

\subsection*{Runtime} \label{sec:app_2}

The runtime of the algorithm critically depends on the demand distribution and size of the street network covered by the demand. The more unique trips are driven, the slower the computation since more different shortest path have to be recomputed in each step. Note, however, that we only recompute shortest paths of cyclists travelling along a bike path that was removed in the current step. Similarly, the size of the covered network in terms of the number of edges directly affects the complexity of the shortest path computation and the number of bike paths that have to be removed in total. Consequently, the larger the covered network, the longer the computational time.

\paragraph{Hamburg} A full run for Hamburg takes approximately 14 minutes with the empirical data (15911 unique trips) and the network as shown in the main manuscript. For the homogeneous demand, the full simulation is slightly slower and takes approximately 15 minutes due to 16002 unique trips of the all-to-all demand with the same network. The difference is small since the empirical demand for Hamburg is already close to all-to-all demand.

\paragraph{Dresden} A full run for Dresden takes approximately 14 minutes with the empirical data (13254 unique trips) and the network as shown in the main manuscript. For the homogeneous demand, the full simulation is significantly slower and takes approximately 29 minutes for the same network. Here, the number of unique trips strongly increases with all-to-all demand to 20022 unique trips.

\paragraph{Hardware} The specified runtimes were achieved on a consumer laptop with an Intel\textsuperscript{\textregistered} Core\texttrademark \:i5-10210U.

\subsection*{Comparison to static percolation approaches} \label{sec:app_3}

Existing static percolation approaches remove bike paths based on a fixed ordering from a snapshot of the street segment importance for the initial, complete bike path network $G_{B}(M)$, starting with the least to most important bike path in the initial setting. Here we chose two different importance measurements: (i) the same importance measurement as explained in the manuscript given by the product $p^0_{ij} \, n_{ij}(M)$ of penalty $p^0_{ij}$ if the street had no bike path and the number of users of that street segment $n_{ij}(M)$, and (ii) solely the number of users of that street segment $n_{ij}(M)$.

For the analysis of the static percolation approach we use the same concept as for our dynamic approach to enable a direct comparison. For each bike path network $G^{\mathrm{static}}_{B}(M')$ we calculate the chosen routes based on our cyclists route choice model and calculate the resulting observables such as the bikeability $b(\lambda)$ and the fraction cycled on a street without a bike path.

\begin{figure}[h]
    \centering
    \includegraphics{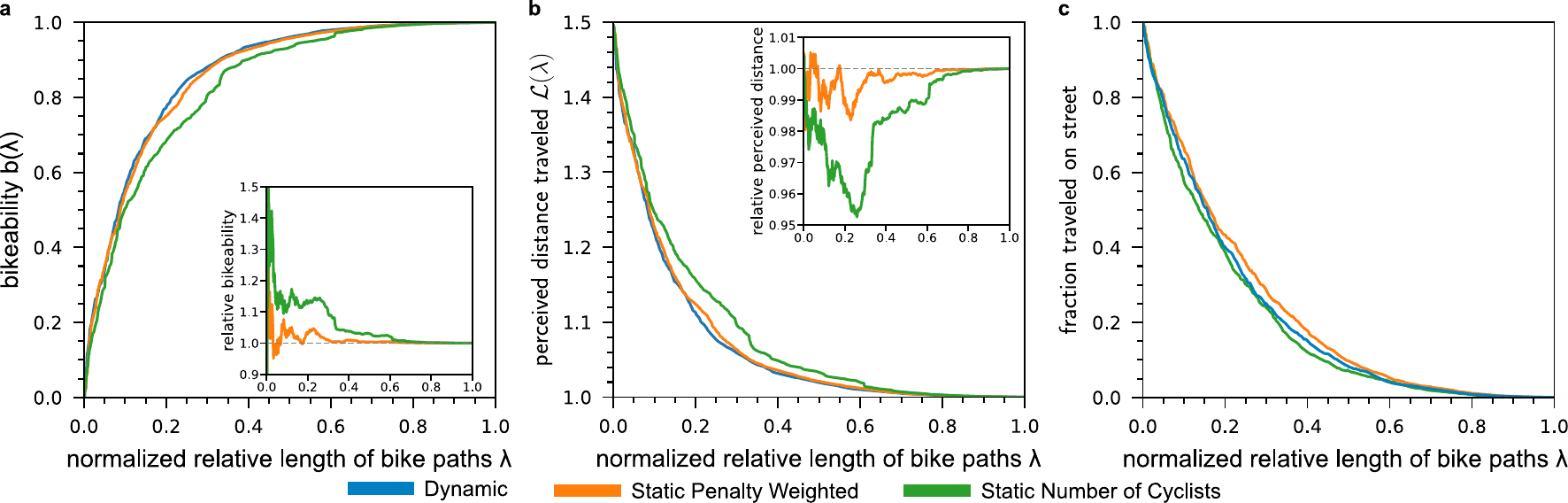}
    \caption{\textbf{Dynamic network generation framework creates highly bikeable networks.}  Comparison between the dynamic framework introduced in the main manuscript (blue) and static approaches based on the same importance measurement (orange) or pure user numbers (green) for the empirical bike-sharing demand in Hamburg (compare Fig.~5 in the main manuscript). The pure user number based measurement is comparable to the approach in \cite{Olmos2020_data}. (a) The dynamic approach generates network with slightly better bikeability $b(\lambda)$ compared to (penalty-weighted) static percolation. These improvements are better visible in terms of the relative bikeability ${b_{dynamic}(\lambda)}/{b_{static}(\lambda)}$ (inset). (b) As for the bikeability, penalty-weighted approaches yield better networks that reduce the perceived distance for cyclists. The inset shows the relative perceived distance  ${\mathcal{L}_{dynamic}(\lambda)}/{\mathcal{L}_{static}(\lambda)}$. (c) All approaches significantly reduce the fraction cycled on streets already with few bike paths. The static approach considering only the number of cyclists using each street segment performs slightly better than the other two, however, no distinction is made between the traffic load of the streets cyclists have to travel on.
    }
    \label{fig_app:dyn_stat}
\end{figure}

The dynamic network generation framework introduced in the manuscript generates only slightly more bikeable networks compared to penalty-weighted static percolation (see Fig.~\ref{fig_app:dyn_stat}). Differences between both approaches would likely increase for more diverse demand distributions with more origin and destination locations and more complex route choice models including additional aspects such as left turns.
Beyond this small increase in network efficiency, and more importantly, our framework opens up a new perspective on bike path network generation by including a flexible model of cyclist route choice behavior. It thus enables fast generation and testing of bike path network structures bridging the gap between static percolation approaches and complex real-world bike path network planning.

\begin{thebibliography}{83}%
\makeatletter
\providecommand \@ifxundefined [1]{%
 \@ifx{#1\undefined}
}%
\providecommand \@ifnum [1]{%
 \ifnum #1\expandafter \@firstoftwo
 \else \expandafter \@secondoftwo
 \fi
}%
\providecommand \@ifx [1]{%
 \ifx #1\expandafter \@firstoftwo
 \else \expandafter \@secondoftwo
 \fi
}%
\providecommand \natexlab [1]{#1}%
\providecommand \enquote  [1]{``#1''}%
\providecommand \bibnamefont  [1]{#1}%
\providecommand \bibfnamefont [1]{#1}%
\providecommand \citenamefont [1]{#1}%
\providecommand \href@noop [0]{\@secondoftwo}%
\providecommand \href [0]{\begingroup \@sanitize@url \@href}%
\providecommand \@href[1]{\@@startlink{#1}\@@href}%
\providecommand \@@href[1]{\endgroup#1\@@endlink}%
\providecommand \@sanitize@url [0]{\catcode `\\12\catcode `\$12\catcode
  `\&12\catcode `\#12\catcode `\^12\catcode `\_12\catcode `\%12\relax}%
\providecommand \@@startlink[1]{}%
\providecommand \@@endlink[0]{}%
\providecommand \url  [0]{\begingroup\@sanitize@url \@url }%
\providecommand \@url [1]{\endgroup\@href {#1}{\urlprefix }}%
\providecommand \urlprefix  [0]{URL }%
\providecommand \Eprint [0]{\href }%
\providecommand \doibase [0]{https://doi.org/}%
\providecommand \selectlanguage [0]{\@gobble}%
\providecommand \bibinfo  [0]{\@secondoftwo}%
\providecommand \bibfield  [0]{\@secondoftwo}%
\providecommand \translation [1]{[#1]}%
\providecommand \BibitemOpen [0]{}%
\providecommand \bibitemStop [0]{}%
\providecommand \bibitemNoStop [0]{.\EOS\space}%
\providecommand \EOS [0]{\spacefactor3000\relax}%
\providecommand \BibitemShut  [1]{\csname bibitem#1\endcsname}%
\let\auto@bib@innerbib\@empty
%</preamble>
\bibitem [{\citenamefont {Banister}(2008)}]{Banister2008}%
  \BibitemOpen
  \bibfield  {author} {\bibinfo {author} {\bibfnamefont {D.}~\bibnamefont
  {Banister}},\ }\bibfield  {title} {\bibinfo {title} {The sustainable mobility
  paradigm},\ }\href {https://doi.org/10.1016/j.tranpol.2007.10.005} {\bibfield
   {journal} {\bibinfo  {journal} {Transp. Policy}\ }\textbf {\bibinfo {volume}
  {15}},\ \bibinfo {pages} {73} (\bibinfo {year} {2008})}\BibitemShut {NoStop}%
\bibitem [{\citenamefont {Mazzoncini}\ \emph {et~al.}(2020)\citenamefont
  {Mazzoncini}, \citenamefont {Somaschini},\ and\ \citenamefont
  {Longo}}]{Mazzoncini2020}%
  \BibitemOpen
  \bibfield  {author} {\bibinfo {author} {\bibfnamefont {R.}~\bibnamefont
  {Mazzoncini}}, \bibinfo {author} {\bibfnamefont {C.}~\bibnamefont
  {Somaschini}},\ and\ \bibinfo {author} {\bibfnamefont {M.}~\bibnamefont
  {Longo}},\ }\bibinfo {title} {The infrastructure for sustainable mobility},\
  in\ \href@noop {} {\emph {\bibinfo {booktitle} {Green Planning for Cities and
  Communities: Novel Incisive Approaches to Sustainability}}},\ \bibinfo
  {editor} {edited by\ \bibinfo {editor} {\bibfnamefont {G.}~\bibnamefont
  {Dall'O'}}}\ (\bibinfo  {publisher} {Springer International Publishing},\
  \bibinfo {address} {Cham},\ \bibinfo {year} {2020})\ pp.\ \bibinfo {pages}
  {255--277}\BibitemShut {NoStop}%
\bibitem [{\citenamefont {Standen}\ \emph {et~al.}(2019)\citenamefont
  {Standen}, \citenamefont {Greaves}, \citenamefont {Collins}, \citenamefont
  {Crane},\ and\ \citenamefont {Rissel}}]{Standen2019}%
  \BibitemOpen
  \bibfield  {author} {\bibinfo {author} {\bibfnamefont {C.}~\bibnamefont
  {Standen}}, \bibinfo {author} {\bibfnamefont {S.}~\bibnamefont {Greaves}},
  \bibinfo {author} {\bibfnamefont {A.~T.}\ \bibnamefont {Collins}}, \bibinfo
  {author} {\bibfnamefont {M.}~\bibnamefont {Crane}},\ and\ \bibinfo {author}
  {\bibfnamefont {C.}~\bibnamefont {Rissel}},\ }\bibfield  {title} {\bibinfo
  {title} {The value of slow travel: Economic appraisal of cycling projects
  using the logsum measure of consumer surplus},\ }\href
  {https://doi.org/10.1016/j.tra.2018.10.015} {\bibfield  {journal} {\bibinfo
  {journal} {Transp. Res. A}\ }\textbf {\bibinfo {volume} {123}},\ \bibinfo
  {pages} {255} (\bibinfo {year} {2019})}\BibitemShut {NoStop}%
\bibitem [{\citenamefont {Creutzig}\ \emph {et~al.}(2016)\citenamefont
  {Creutzig}, \citenamefont {Agoston}, \citenamefont {Minx}, \citenamefont
  {Canadell}, \citenamefont {Andrew}, \citenamefont {Qu{\'{e}}r{\'{e}}},
  \citenamefont {Peters}, \citenamefont {Sharifi}, \citenamefont {Yamagata},\
  and\ \citenamefont {Dhakal}}]{Creutzig2016}%
  \BibitemOpen
  \bibfield  {author} {\bibinfo {author} {\bibfnamefont {F.}~\bibnamefont
  {Creutzig}}, \bibinfo {author} {\bibfnamefont {P.}~\bibnamefont {Agoston}},
  \bibinfo {author} {\bibfnamefont {J.~C.}\ \bibnamefont {Minx}}, \bibinfo
  {author} {\bibfnamefont {J.~G.}\ \bibnamefont {Canadell}}, \bibinfo {author}
  {\bibfnamefont {R.~M.}\ \bibnamefont {Andrew}}, \bibinfo {author}
  {\bibfnamefont {C.~L.}\ \bibnamefont {Qu{\'{e}}r{\'{e}}}}, \bibinfo {author}
  {\bibfnamefont {G.~P.}\ \bibnamefont {Peters}}, \bibinfo {author}
  {\bibfnamefont {A.}~\bibnamefont {Sharifi}}, \bibinfo {author} {\bibfnamefont
  {Y.}~\bibnamefont {Yamagata}},\ and\ \bibinfo {author} {\bibfnamefont
  {S.}~\bibnamefont {Dhakal}},\ }\bibfield  {title} {\bibinfo {title} {Urban
  infrastructure choices structure climate solutions},\ }\href
  {https://doi.org/10.1038/nclimate3169} {\bibfield  {journal} {\bibinfo
  {journal} {Nat. Clim. Change}\ }\textbf {\bibinfo {volume} {6}},\ \bibinfo
  {pages} {1054} (\bibinfo {year} {2016})}\BibitemShut {NoStop}%
\bibitem [{\citenamefont {Creutzig}\ \emph {et~al.}(2015)\citenamefont
  {Creutzig}, \citenamefont {Jochem}, \citenamefont {Edelenbosch},
  \citenamefont {Mattauch}, \citenamefont {Vuuren}, \citenamefont {McCollum},\
  and\ \citenamefont {Minx}}]{Creutzig2015}%
  \BibitemOpen
  \bibfield  {author} {\bibinfo {author} {\bibfnamefont {F.}~\bibnamefont
  {Creutzig}}, \bibinfo {author} {\bibfnamefont {P.}~\bibnamefont {Jochem}},
  \bibinfo {author} {\bibfnamefont {O.~Y.}\ \bibnamefont {Edelenbosch}},
  \bibinfo {author} {\bibfnamefont {L.}~\bibnamefont {Mattauch}}, \bibinfo
  {author} {\bibfnamefont {D.~P.~v.}\ \bibnamefont {Vuuren}}, \bibinfo {author}
  {\bibfnamefont {D.}~\bibnamefont {McCollum}},\ and\ \bibinfo {author}
  {\bibfnamefont {J.}~\bibnamefont {Minx}},\ }\bibfield  {title} {\bibinfo
  {title} {Transport: A roadblock to climate change mitigation?},\ }\href
  {https://doi.org/10.1126/science.aac8033} {\bibfield  {journal} {\bibinfo
  {journal} {Science}\ }\textbf {\bibinfo {volume} {350}},\ \bibinfo {pages}
  {911} (\bibinfo {year} {2015})}\BibitemShut {NoStop}%
\bibitem [{\citenamefont {Sims}\ \emph {et~al.}(2014)\citenamefont {Sims},
  \citenamefont {Schaeffer}, \citenamefont {Creutzig}, \citenamefont
  {Cruz-Núñez}, \citenamefont {D’Agosto}, \citenamefont {Dimitriu},
  \citenamefont {Meza}, \citenamefont {Fulton}, \citenamefont {Kobayashi},
  \citenamefont {Lah}, \citenamefont {McKinnon}, \citenamefont {Newman},
  \citenamefont {Ouyang}, \citenamefont {Schauer}, \citenamefont {Sperling},\
  and\ \citenamefont {Tiwari}}]{IPCC2014}%
  \BibitemOpen
  \bibfield  {author} {\bibinfo {author} {\bibfnamefont {R.}~\bibnamefont
  {Sims}}, \bibinfo {author} {\bibfnamefont {R.}~\bibnamefont {Schaeffer}},
  \bibinfo {author} {\bibfnamefont {F.}~\bibnamefont {Creutzig}}, \bibinfo
  {author} {\bibfnamefont {X.}~\bibnamefont {Cruz-Núñez}}, \bibinfo {author}
  {\bibfnamefont {M.}~\bibnamefont {D’Agosto}}, \bibinfo {author}
  {\bibfnamefont {D.}~\bibnamefont {Dimitriu}}, \bibinfo {author}
  {\bibfnamefont {M.~J.~F.}\ \bibnamefont {Meza}}, \bibinfo {author}
  {\bibfnamefont {L.}~\bibnamefont {Fulton}}, \bibinfo {author} {\bibfnamefont
  {S.}~\bibnamefont {Kobayashi}}, \bibinfo {author} {\bibfnamefont
  {O.}~\bibnamefont {Lah}}, \bibinfo {author} {\bibfnamefont {A.}~\bibnamefont
  {McKinnon}}, \bibinfo {author} {\bibfnamefont {P.}~\bibnamefont {Newman}},
  \bibinfo {author} {\bibfnamefont {M.}~\bibnamefont {Ouyang}}, \bibinfo
  {author} {\bibfnamefont {J.~J.}\ \bibnamefont {Schauer}}, \bibinfo {author}
  {\bibfnamefont {D.}~\bibnamefont {Sperling}},\ and\ \bibinfo {author}
  {\bibfnamefont {G.}~\bibnamefont {Tiwari}},\ }\bibfield  {title} {\bibinfo
  {title} {{Transport. Contribution of Working Group III to the Fifth
  Assessment Report of the Intergovernmental Panel on Climate Change}},\ }in\
  \href@noop {} {\emph {\bibinfo {booktitle} {Climate Change 2014: Mitigation
  of Climate Change. Contribution of Working Group III to the Fifth Assessment
  Report of the Intergovern-mental Panel on Climate Change}}},\ \bibinfo
  {editor} {edited by\ \bibinfo {editor} {\bibfnamefont {O.}~\bibnamefont
  {Edenhofer}}, \bibinfo {editor} {\bibfnamefont {R.}~\bibnamefont
  {Pichs-Madruga}}, \bibinfo {editor} {\bibfnamefont {Y.}~\bibnamefont
  {Sokona}}, \bibinfo {editor} {\bibfnamefont {E.}~\bibnamefont {Farahani}},
  \bibinfo {editor} {\bibfnamefont {S.}~\bibnamefont {Kadner}}, \bibinfo
  {editor} {\bibfnamefont {K.}~\bibnamefont {Seyboth}}, \bibinfo {editor}
  {\bibfnamefont {A.}~\bibnamefont {Adler}}, \bibinfo {editor} {\bibfnamefont
  {I.}~\bibnamefont {Baum}}, \bibinfo {editor} {\bibfnamefont {S.}~\bibnamefont
  {Brunner}}, \bibinfo {editor} {\bibfnamefont {P.}~\bibnamefont {Eickemeier}},
  \bibinfo {editor} {\bibfnamefont {B.}~\bibnamefont {Kriemann}}, \bibinfo
  {editor} {\bibfnamefont {J.}~\bibnamefont {Savolainen}}, \bibinfo {editor}
  {\bibfnamefont {S.}~\bibnamefont {Schlömer}}, \bibinfo {editor}
  {\bibfnamefont {C.}~\bibnamefont {von Stechow}}, \bibinfo {editor}
  {\bibfnamefont {T.}~\bibnamefont {Zwickel}},\ and\ \bibinfo {editor}
  {\bibfnamefont {J.~C.}\ \bibnamefont {Minx}}}\ (\bibinfo  {publisher}
  {Cambridge University Press},\ \bibinfo {address} {Cambridge, United Kingdom
  and New York, NY, USA},\ \bibinfo {year} {2014})\ pp.\ \bibinfo {pages}
  {599--670}\BibitemShut {NoStop}%
\bibitem [{\citenamefont {Li}\ \emph {et~al.}(2022)\citenamefont {Li},
  \citenamefont {Luo}, \citenamefont {Shang}, \citenamefont {Lv}, \citenamefont
  {Fan}, \citenamefont {Lu}, \citenamefont {Pan}, \citenamefont {Tian},\ and\
  \citenamefont {Stanley}}]{Li2022_scaling}%
  \BibitemOpen
  \bibfield  {author} {\bibinfo {author} {\bibfnamefont {R.}~\bibnamefont
  {Li}}, \bibinfo {author} {\bibfnamefont {A.}~\bibnamefont {Luo}}, \bibinfo
  {author} {\bibfnamefont {F.}~\bibnamefont {Shang}}, \bibinfo {author}
  {\bibfnamefont {L.}~\bibnamefont {Lv}}, \bibinfo {author} {\bibfnamefont
  {J.}~\bibnamefont {Fan}}, \bibinfo {author} {\bibfnamefont {G.}~\bibnamefont
  {Lu}}, \bibinfo {author} {\bibfnamefont {L.}~\bibnamefont {Pan}}, \bibinfo
  {author} {\bibfnamefont {L.}~\bibnamefont {Tian}},\ and\ \bibinfo {author}
  {\bibfnamefont {H.~E.}\ \bibnamefont {Stanley}},\ }\bibfield  {title}
  {\bibinfo {title} {Emergence of scaling in dockless bike-sharing systems},\
  }\href@noop {} {\bibfield  {journal} {\bibinfo  {journal} {arXiv preprint}\ }
  (\bibinfo {year} {2022})},\ \Eprint {https://arxiv.org/abs/2202.06352}
  {arXiv:2202.06352} \BibitemShut {NoStop}%
\bibitem [{\citenamefont {Rhoads}\ \emph {et~al.}(2021)\citenamefont {Rhoads},
  \citenamefont {Sol{\'{e}}-Ribalta}, \citenamefont {Gonz{\'{a}}lez},\ and\
  \citenamefont {Borge-Holthoefer}}]{Rhoads2021}%
  \BibitemOpen
  \bibfield  {author} {\bibinfo {author} {\bibfnamefont {D.}~\bibnamefont
  {Rhoads}}, \bibinfo {author} {\bibfnamefont {A.}~\bibnamefont
  {Sol{\'{e}}-Ribalta}}, \bibinfo {author} {\bibfnamefont {M.~C.}\ \bibnamefont
  {Gonz{\'{a}}lez}},\ and\ \bibinfo {author} {\bibfnamefont {J.}~\bibnamefont
  {Borge-Holthoefer}},\ }\bibfield  {title} {\bibinfo {title} {A sustainable
  strategy for open streets in (post)pandemic cities},\ }\href {https://doi.org/10.1038/s42005-021-00688-z} {\bibfield  {journal}
  {\bibinfo  {journal} {Commun. Phys.}\ }\textbf {\bibinfo {volume} {4}},\ (\bibinfo {year} {2021})}\BibitemShut {NoStop}%
\bibitem [{\citenamefont {Vandy}(2020)}]{BBC2020}%
  \BibitemOpen
  \bibfield  {author} {\bibinfo {author} {\bibfnamefont {K.}~\bibnamefont
  {Vandy}},\ }\href@noop {} {\bibinfo {title} {Coronavirus: How pandemic
  sparked european cycling revolution}},\ \bibinfo {howpublished} {BBC}
  (\bibinfo {year} {2020}),\ \bibinfo {note}
  {\url{https://www.bbc.com/news/world-europe-54353914} (accessed on
  2021-11-24)}\BibitemShut {NoStop}%
\bibitem [{\citenamefont {Schwedhelm}\ \emph {et~al.}(2020)\citenamefont
  {Schwedhelm}, \citenamefont {Li}, \citenamefont {Harms},\ and\ \citenamefont
  {Adriazola-Steil}}]{WRI2020}%
  \BibitemOpen
  \bibfield  {author} {\bibinfo {author} {\bibfnamefont {A.}~\bibnamefont
  {Schwedhelm}}, \bibinfo {author} {\bibfnamefont {W.}~\bibnamefont {Li}},
  \bibinfo {author} {\bibfnamefont {L.}~\bibnamefont {Harms}},\ and\ \bibinfo
  {author} {\bibfnamefont {C.}~\bibnamefont {Adriazola-Steil}},\ }\href@noop {}
  {\bibinfo {title} {{Cycling during COVID-19}}},\ \bibinfo {howpublished}
  {{World Resources Institute}} (\bibinfo {year} {2020}),\ \bibinfo {note}
  {\url{https://www.wri.org/blog/2020/04/coronavirus-biking-critical-in-cities}
  (accessed on 2021-11-24)}\BibitemShut {NoStop}%
\bibitem [{\citenamefont {Goetsch}\ and\ \citenamefont
  {Quiros}(2020)}]{WorldBank2020}%
  \BibitemOpen
  \bibfield  {author} {\bibinfo {author} {\bibfnamefont {H.}~\bibnamefont
  {Goetsch}}\ and\ \bibinfo {author} {\bibfnamefont {T.~P.}\ \bibnamefont
  {Quiros}},\ }\href@noop {} {\bibinfo {title} {{COVID-19 creates new momentum
  for cycling and walking. We can't let it go to waste!}}},\ \bibinfo
  {howpublished} {{World Bank}} (\bibinfo {year} {2020}),\ \bibinfo {note}
  {\url{https://blogs.worldbank.org/transport/covid-19-creates-new-momentum-cycling-and-walking-we-cant-let-it-go-waste}
  (accessed on 2021-11-24)}\BibitemShut {NoStop}%
\bibitem [{\citenamefont {Oltermann}(2020)}]{Guardian2020}%
  \BibitemOpen
  \bibfield  {author} {\bibinfo {author} {\bibfnamefont {P.}~\bibnamefont
  {Oltermann}},\ }\href@noop {} {\bibinfo {title} {Pop-up bike lanes help with
  coronavirus physical distancing in germany}},\ \bibinfo {howpublished} {{The
  Guardian}} (\bibinfo {year} {2020}),\ \bibinfo {note}
  {\url{https://www.theguardian.com/world/2020/apr/13/pop-up-bike-lanes-help-with-coronavirus-social-distancing-in-germany}
  (accessed on 2021-11-24)}\BibitemShut {NoStop}%
\bibitem [{\citenamefont {Sadik-Khan}\ and\ \citenamefont
  {Solomonow}(2021)}]{Guardian2021}%
  \BibitemOpen
  \bibfield  {author} {\bibinfo {author} {\bibfnamefont {J.}~\bibnamefont
  {Sadik-Khan}}\ and\ \bibinfo {author} {\bibfnamefont {S.}~\bibnamefont
  {Solomonow}},\ }\href@noop {} {\bibinfo {title} {The bikelash paradox: How
  cycle lanes enrage some but win votes}},\ \bibinfo {howpublished} {{The
  Guardian}} (\bibinfo {year} {2021}),\ \bibinfo {note}
  {\url{https://www.theguardian.com/environment/bike-blog/2021/oct/29/the-bikelash-paradox-how-cycle-lanes-enrage-some-but-win-votes}
  (accessed on 2021-11-24)}\BibitemShut {NoStop}%
\bibitem [{\citenamefont {Jackson}(2010)}]{Jackson2010_social}%
  \BibitemOpen
  \bibfield  {author} {\bibinfo {author} {\bibfnamefont {M.~O.}\ \bibnamefont
  {Jackson}},\ }\href@noop {} {\emph {\bibinfo {title} {Social and Economic
  Networks}}}\ (\bibinfo  {publisher} {Princeton University Press},\ \bibinfo
  {address} {Princeton},\ \bibinfo {year} {2010})\BibitemShut {NoStop}%
\bibitem [{\citenamefont {Schr\"oder}\ \emph {et~al.}(2018)\citenamefont
  {Schr\"oder}, \citenamefont {Nagler}, \citenamefont {Timme},\ and\
  \citenamefont {Witthaut}}]{Schoreder2018_hysteretic}%
  \BibitemOpen
  \bibfield  {author} {\bibinfo {author} {\bibfnamefont {M.}~\bibnamefont
  {Schr\"oder}}, \bibinfo {author} {\bibfnamefont {J.}~\bibnamefont {Nagler}},
  \bibinfo {author} {\bibfnamefont {M.}~\bibnamefont {Timme}},\ and\ \bibinfo
  {author} {\bibfnamefont {D.}~\bibnamefont {Witthaut}},\ }\bibfield  {title}
  {\bibinfo {title} {Hysteretic percolation from locally optimal individual
  decisions},\ }\href {https://doi.org/10.1103/PhysRevLett.120.248302}
  {\bibfield  {journal} {\bibinfo  {journal} {Phys. Rev. Lett.}\ }\textbf
  {\bibinfo {volume} {120}},\ \bibinfo {pages} {248302} (\bibinfo {year}
  {2018})}\BibitemShut {NoStop}%
\bibitem [{\citenamefont {Aldous}\ and\ \citenamefont
  {Barthelemy}(2019)}]{Aldous2019}%
  \BibitemOpen
  \bibfield  {author} {\bibinfo {author} {\bibfnamefont {D.}~\bibnamefont
  {Aldous}}\ and\ \bibinfo {author} {\bibfnamefont {M.}~\bibnamefont
  {Barthelemy}},\ }\bibfield  {title} {\bibinfo {title} {Optimal geometry of
  transportation networks},\ }\href
  {https://doi.org/10.1103/PhysRevE.99.052303} {\bibfield  {journal} {\bibinfo
  {journal} {Phys. Rev. E}\ }\textbf {\bibinfo {volume} {99}},\ \bibinfo
  {pages} {052303} (\bibinfo {year} {2019})}\BibitemShut {NoStop}%
\bibitem [{\citenamefont {Frangopol}\ and\ \citenamefont
  {Liu}(2007)}]{Frangopol2007}%
  \BibitemOpen
  \bibfield  {author} {\bibinfo {author} {\bibfnamefont {D.~M.}\ \bibnamefont
  {Frangopol}}\ and\ \bibinfo {author} {\bibfnamefont {M.}~\bibnamefont
  {Liu}},\ }\bibfield  {title} {\bibinfo {title} {Maintenance and management of
  civil infrastructure based on condition, safety, optimization, and life-cycle
  cost},\ }\href {https://doi.org/10.1080/15732470500253164} {\bibfield
  {journal} {\bibinfo  {journal} {Struct. Infrastruct. E}\ }\textbf {\bibinfo
  {volume} {3}},\ \bibinfo {pages} {29} (\bibinfo {year} {2007})}\BibitemShut
  {NoStop}%
\bibitem [{\citenamefont {Tero}\ \emph {et~al.}(2010)\citenamefont {Tero},
  \citenamefont {Takagi}, \citenamefont {Saigusa}, \citenamefont {Ito},
  \citenamefont {Bebber}, \citenamefont {Fricker}, \citenamefont {Yumiki},
  \citenamefont {Kobayashi},\ and\ \citenamefont {Nakagaki}}]{Tero2010_rules}%
  \BibitemOpen
  \bibfield  {author} {\bibinfo {author} {\bibfnamefont {A.}~\bibnamefont
  {Tero}}, \bibinfo {author} {\bibfnamefont {S.}~\bibnamefont {Takagi}},
  \bibinfo {author} {\bibfnamefont {T.}~\bibnamefont {Saigusa}}, \bibinfo
  {author} {\bibfnamefont {K.}~\bibnamefont {Ito}}, \bibinfo {author}
  {\bibfnamefont {D.~P.}\ \bibnamefont {Bebber}}, \bibinfo {author}
  {\bibfnamefont {M.~D.}\ \bibnamefont {Fricker}}, \bibinfo {author}
  {\bibfnamefont {K.}~\bibnamefont {Yumiki}}, \bibinfo {author} {\bibfnamefont
  {R.}~\bibnamefont {Kobayashi}},\ and\ \bibinfo {author} {\bibfnamefont
  {T.}~\bibnamefont {Nakagaki}},\ }\bibfield  {title} {\bibinfo {title} {Rules
  for biologically inspired adaptive network design},\ }\href
  {https://doi.org/10.1126/science.1177894} {\bibfield  {journal} {\bibinfo
  {journal} {Science}\ }\textbf {\bibinfo {volume} {327}},\ \bibinfo {pages}
  {439} (\bibinfo {year} {2010})}\BibitemShut {NoStop}%
\bibitem [{\citenamefont {Katifori}\ \emph {et~al.}(2010)\citenamefont
  {Katifori}, \citenamefont {Sz\"oll\ifmmode~\mbox{\H{o}}\else \H{o}\fi{}si},\
  and\ \citenamefont {Magnasco}}]{Katifori2010_damage}%
  \BibitemOpen
  \bibfield  {author} {\bibinfo {author} {\bibfnamefont {E.}~\bibnamefont
  {Katifori}}, \bibinfo {author} {\bibfnamefont {G.~J.}\ \bibnamefont
  {Sz\"oll\ifmmode~\mbox{\H{o}}\else \H{o}\fi{}si}},\ and\ \bibinfo {author}
  {\bibfnamefont {M.~O.}\ \bibnamefont {Magnasco}},\ }\bibfield  {title}
  {\bibinfo {title} {Damage and fluctuations induce loops in optimal transport
  networks},\ }\href {https://doi.org/10.1103/PhysRevLett.104.048704}
  {\bibfield  {journal} {\bibinfo  {journal} {Phys. Rev. Lett.}\ }\textbf
  {\bibinfo {volume} {104}},\ \bibinfo {pages} {048704} (\bibinfo {year}
  {2010})}\BibitemShut {NoStop}%
\bibitem [{\citenamefont {Ronellenfitsch}\ and\ \citenamefont
  {Katifori}(2016)}]{Ronellenfitsch2016_global}%
  \BibitemOpen
  \bibfield  {author} {\bibinfo {author} {\bibfnamefont {H.}~\bibnamefont
  {Ronellenfitsch}}\ and\ \bibinfo {author} {\bibfnamefont {E.}~\bibnamefont
  {Katifori}},\ }\bibfield  {title} {\bibinfo {title} {Global optimization,
  local adaptation, and the role of growth in distribution networks},\ }\href
  {https://doi.org/10.1103/PhysRevLett.117.138301} {\bibfield  {journal}
  {\bibinfo  {journal} {Phys. Rev. Lett.}\ }\textbf {\bibinfo {volume} {117}},\
  \bibinfo {pages} {138301} (\bibinfo {year} {2016})}\BibitemShut {NoStop}%
\bibitem [{\citenamefont {Karschau}\ \emph {et~al.}(2020)\citenamefont
  {Karschau}, \citenamefont {Scholich}, \citenamefont {Wise}, \citenamefont
  {Morales-Navarrete}, \citenamefont {Kalaidzidis}, \citenamefont {Zerial},\
  and\ \citenamefont {Friedrich}}]{Karschau2020_resilience}%
  \BibitemOpen
  \bibfield  {author} {\bibinfo {author} {\bibfnamefont {J.}~\bibnamefont
  {Karschau}}, \bibinfo {author} {\bibfnamefont {A.}~\bibnamefont {Scholich}},
  \bibinfo {author} {\bibfnamefont {J.}~\bibnamefont {Wise}}, \bibinfo {author}
  {\bibfnamefont {H.}~\bibnamefont {Morales-Navarrete}}, \bibinfo {author}
  {\bibfnamefont {Y.}~\bibnamefont {Kalaidzidis}}, \bibinfo {author}
  {\bibfnamefont {M.}~\bibnamefont {Zerial}},\ and\ \bibinfo {author}
  {\bibfnamefont {B.~M.}\ \bibnamefont {Friedrich}},\ }\bibfield  {title}
  {\bibinfo {title} {Resilience of three-dimensional sinusoidal networks in
  liver tissue},\ }\href {https://doi.org/10.1371/journal.pcbi.1007965}
  {\bibfield  {journal} {\bibinfo  {journal} {PLOS Comput. Biol.}\ }\textbf
  {\bibinfo {volume} {16}},\ \bibinfo {pages} {1} (\bibinfo {year}
  {2020})}\BibitemShut {NoStop}%
\bibitem [{\citenamefont {Kleinberg}(2000)}]{Kleinberg2000_navigation}%
  \BibitemOpen
  \bibfield  {author} {\bibinfo {author} {\bibfnamefont {J.~M.}\ \bibnamefont
  {Kleinberg}},\ }\bibfield  {title} {\bibinfo {title} {Navigation in a small
  world},\ }\href {https://doi.org/10.1038/35022643} {\bibfield  {journal}
  {\bibinfo  {journal} {Nature}\ }\textbf {\bibinfo {volume} {406}},\ \bibinfo
  {pages} {845} (\bibinfo {year} {2000})}\BibitemShut {NoStop}%
\bibitem [{\citenamefont {Cohen}\ \emph {et~al.}(2000)\citenamefont {Cohen},
  \citenamefont {Erez}, \citenamefont {ben Avraham},\ and\ \citenamefont
  {Havlin}}]{cohen00_resilience}%
  \BibitemOpen
  \bibfield  {author} {\bibinfo {author} {\bibfnamefont {R.}~\bibnamefont
  {Cohen}}, \bibinfo {author} {\bibfnamefont {K.}~\bibnamefont {Erez}},
  \bibinfo {author} {\bibfnamefont {D.}~\bibnamefont {ben Avraham}},\ and\
  \bibinfo {author} {\bibfnamefont {S.}~\bibnamefont {Havlin}},\ }\bibfield
  {title} {\bibinfo {title} {Resilience of the internet to random breakdowns},\
  }\href {https://doi.org/10.1103/PhysRevLett.85.4626} {\bibfield  {journal}
  {\bibinfo  {journal} {Phys. Rev. Lett.}\ }\textbf {\bibinfo {volume} {85}},\
  \bibinfo {pages} {4626} (\bibinfo {year} {2000})}\BibitemShut {NoStop}%
\bibitem [{\citenamefont {Molkenthin}\ \emph {et~al.}(2018)\citenamefont
  {Molkenthin}, \citenamefont {Schr\"oder},\ and\ \citenamefont
  {Timme}}]{Molkenthin2018Adhesion}%
  \BibitemOpen
  \bibfield  {author} {\bibinfo {author} {\bibfnamefont {N.}~\bibnamefont
  {Molkenthin}}, \bibinfo {author} {\bibfnamefont {M.}~\bibnamefont
  {Schr\"oder}},\ and\ \bibinfo {author} {\bibfnamefont {M.}~\bibnamefont
  {Timme}},\ }\bibfield  {title} {\bibinfo {title} {Adhesion-induced
  discontinuous transitions and classifying social networks},\ }\href
  {https://doi.org/10.1103/PhysRevLett.121.138301} {\bibfield  {journal}
  {\bibinfo  {journal} {Phys. Rev. Lett.}\ }\textbf {\bibinfo {volume} {121}},\
  \bibinfo {pages} {138301} (\bibinfo {year} {2018})}\BibitemShut {NoStop}%
\bibitem [{\citenamefont {Gastner}\ and\ \citenamefont
  {Newman}(2006)}]{Gastner2006_optimal}%
  \BibitemOpen
  \bibfield  {author} {\bibinfo {author} {\bibfnamefont {M.~T.}\ \bibnamefont
  {Gastner}}\ and\ \bibinfo {author} {\bibfnamefont {M.~E.~J.}\ \bibnamefont
  {Newman}},\ }\bibfield  {title} {\bibinfo {title} {Optimal design of spatial
  distribution networks},\ }\href {https://doi.org/10.1103/PhysRevE.74.016117}
  {\bibfield  {journal} {\bibinfo  {journal} {Phys. Rev. E}\ }\textbf {\bibinfo
  {volume} {74}},\ \bibinfo {pages} {016117} (\bibinfo {year}
  {2006})}\BibitemShut {NoStop}%
\bibitem [{\citenamefont {Barth{\'e}lemy}\ and\ \citenamefont
  {Flammini}(2006)}]{Barthelemy2006_optimal}%
  \BibitemOpen
  \bibfield  {author} {\bibinfo {author} {\bibfnamefont {M.}~\bibnamefont
  {Barth{\'e}lemy}}\ and\ \bibinfo {author} {\bibfnamefont {A.}~\bibnamefont
  {Flammini}},\ }\bibfield  {title} {\bibinfo {title} {Optimal traffic
  networks},\ }\href {https://doi.org/10.1088/1742-5468/2006/07/L07002}
  {\bibfield  {journal} {\bibinfo  {journal} {J. Stat. Mech. Theory Exp.}\
  }\textbf {\bibinfo {volume} {2006}},\ \bibinfo {pages} {L07002} (\bibinfo
  {year} {2006})}\BibitemShut {NoStop}%
\bibitem [{\citenamefont {Verma}\ \emph {et~al.}(2016)\citenamefont {Verma},
  \citenamefont {Russmann}, \citenamefont {Ara{\'u}jo}, \citenamefont
  {Nagler},\ and\ \citenamefont {Herrmann}}]{Verma2016_emergence}%
  \BibitemOpen
  \bibfield  {author} {\bibinfo {author} {\bibfnamefont {T.}~\bibnamefont
  {Verma}}, \bibinfo {author} {\bibfnamefont {F.}~\bibnamefont {Russmann}},
  \bibinfo {author} {\bibfnamefont {N.~A.}\ \bibnamefont {Ara{\'u}jo}},
  \bibinfo {author} {\bibfnamefont {J.}~\bibnamefont {Nagler}},\ and\ \bibinfo
  {author} {\bibfnamefont {H.~J.}\ \bibnamefont {Herrmann}},\ }\bibfield
  {title} {\bibinfo {title} {Emergence of core--peripheries in networks},\
  }\href {https://doi.org/10.1038/ncomms10441} {\bibfield  {journal} {\bibinfo
  {journal} {Nat. Commun.}\ }\textbf {\bibinfo {volume} {7}},\ \bibinfo {pages}
  {10441} (\bibinfo {year} {2016})}\BibitemShut {NoStop}%
\bibitem [{\citenamefont {Cardillo}\ \emph {et~al.}(2013)\citenamefont
  {Cardillo}, \citenamefont {G{\'o}mez-Gardenes}, \citenamefont {Zanin},
  \citenamefont {Romance}, \citenamefont {Papo}, \citenamefont {Del~Pozo},\
  and\ \citenamefont {Boccaletti}}]{Cardillo2013_emergence}%
  \BibitemOpen
  \bibfield  {author} {\bibinfo {author} {\bibfnamefont {A.}~\bibnamefont
  {Cardillo}}, \bibinfo {author} {\bibfnamefont {J.}~\bibnamefont
  {G{\'o}mez-Gardenes}}, \bibinfo {author} {\bibfnamefont {M.}~\bibnamefont
  {Zanin}}, \bibinfo {author} {\bibfnamefont {M.}~\bibnamefont {Romance}},
  \bibinfo {author} {\bibfnamefont {D.}~\bibnamefont {Papo}}, \bibinfo {author}
  {\bibfnamefont {F.}~\bibnamefont {Del~Pozo}},\ and\ \bibinfo {author}
  {\bibfnamefont {S.}~\bibnamefont {Boccaletti}},\ }\bibfield  {title}
  {\bibinfo {title} {Emergence of network features from multiplexity},\ }\href
  {https://doi.org/10.1038/srep01344} {\bibfield  {journal} {\bibinfo
  {journal} {Sci. Rep.}\ }\textbf {\bibinfo {volume} {3}},\ \bibinfo {pages}
  {1344} (\bibinfo {year} {2013})}\BibitemShut {NoStop}%
\bibitem [{\citenamefont {Barth{\'e}lemy}(2011)}]{Barthelemy2011_spatial}%
  \BibitemOpen
  \bibfield  {author} {\bibinfo {author} {\bibfnamefont {M.}~\bibnamefont
  {Barth{\'e}lemy}},\ }\bibfield  {title} {\bibinfo {title} {Spatial
  networks},\ }\href {https://doi.org/10.1016/j.physrep.2010.11.002} {\bibfield
   {journal} {\bibinfo  {journal} {Phys. Rep.}\ }\textbf {\bibinfo {volume}
  {499}},\ \bibinfo {pages} {1} (\bibinfo {year} {2011})}\BibitemShut {NoStop}%
\bibitem [{\citenamefont {Scellato}\ \emph {et~al.}(2006)\citenamefont
  {Scellato}, \citenamefont {Cardillo}, \citenamefont {Latora},\ and\
  \citenamefont {Porta}}]{Scellato2006_backbone}%
  \BibitemOpen
  \bibfield  {author} {\bibinfo {author} {\bibfnamefont {S.}~\bibnamefont
  {Scellato}}, \bibinfo {author} {\bibfnamefont {A.}~\bibnamefont {Cardillo}},
  \bibinfo {author} {\bibfnamefont {V.}~\bibnamefont {Latora}},\ and\ \bibinfo
  {author} {\bibfnamefont {S.}~\bibnamefont {Porta}},\ }\bibfield  {title}
  {\bibinfo {title} {The backbone of a city},\ }\href
  {https://doi.org/10.1140/epjb/e2006-00066-4} {\bibfield  {journal} {\bibinfo
  {journal} {Eur. Phys. J. B}\ }\textbf {\bibinfo {volume} {50}},\ \bibinfo
  {pages} {221} (\bibinfo {year} {2006})}\BibitemShut {NoStop}%
\bibitem [{\citenamefont {Barth\'elemy}\ and\ \citenamefont
  {Flammini}(2008)}]{Barethelemy2008_modeling}%
  \BibitemOpen
  \bibfield  {author} {\bibinfo {author} {\bibfnamefont {M.}~\bibnamefont
  {Barth\'elemy}}\ and\ \bibinfo {author} {\bibfnamefont {A.}~\bibnamefont
  {Flammini}},\ }\bibfield  {title} {\bibinfo {title} {Modeling urban street
  patterns},\ }\href {https://doi.org/10.1103/PhysRevLett.100.138702}
  {\bibfield  {journal} {\bibinfo  {journal} {Phys. Rev. Lett.}\ }\textbf
  {\bibinfo {volume} {100}},\ \bibinfo {pages} {138702} (\bibinfo {year}
  {2008})}\BibitemShut {NoStop}%
\bibitem [{\citenamefont {Kirkley}\ \emph {et~al.}(2018)\citenamefont
  {Kirkley}, \citenamefont {Barbosa}, \citenamefont {Barthelemy},\ and\
  \citenamefont {Ghoshal}}]{Kirkley2018_betweenness}%
  \BibitemOpen
  \bibfield  {author} {\bibinfo {author} {\bibfnamefont {A.}~\bibnamefont
  {Kirkley}}, \bibinfo {author} {\bibfnamefont {H.}~\bibnamefont {Barbosa}},
  \bibinfo {author} {\bibfnamefont {M.}~\bibnamefont {Barthelemy}},\ and\
  \bibinfo {author} {\bibfnamefont {G.}~\bibnamefont {Ghoshal}},\ }\bibfield
  {title} {\bibinfo {title} {From the betweenness centrality in street networks
  to structural invariants in random planar graphs},\ }\href
  {https://doi.org/10.1038/s41467-018-04978-z} {\bibfield  {journal} {\bibinfo
  {journal} {Nat. Commun.}\ }\textbf {\bibinfo {volume} {9}},\ \bibinfo {pages}
  {2501} (\bibinfo {year} {2018})}\BibitemShut {NoStop}%
\bibitem [{\citenamefont {Duthie}\ and\ \citenamefont
  {Unnikrishnan}(2014)}]{Duthie2014_optimization}%
  \BibitemOpen
  \bibfield  {author} {\bibinfo {author} {\bibfnamefont {J.}~\bibnamefont
  {Duthie}}\ and\ \bibinfo {author} {\bibfnamefont {A.}~\bibnamefont
  {Unnikrishnan}},\ }\bibfield  {title} {\bibinfo {title} {Optimization
  framework for bicycle network design},\ }\href
  {https://doi.org/10.1061/(ASCE)TE.1943-5436.0000690} {\bibfield  {journal}
  {\bibinfo  {journal} {J. Transp. Eng. Part A Syst.}\ }\textbf {\bibinfo
  {volume} {140}},\ \bibinfo {pages} {04014028} (\bibinfo {year}
  {2014})}\BibitemShut {NoStop}%
\bibitem [{\citenamefont {B{\'{i}}l}\ \emph {et~al.}(2015)\citenamefont
  {B{\'{i}}l}, \citenamefont {Andr{\'{a}}{\v{s}}ik},\ and\ \citenamefont
  {Kube{\v{c}}ek}}]{Bil2015}%
  \BibitemOpen
  \bibfield  {author} {\bibinfo {author} {\bibfnamefont {M.}~\bibnamefont
  {B{\'{i}}l}}, \bibinfo {author} {\bibfnamefont {R.}~\bibnamefont
  {Andr{\'{a}}{\v{s}}ik}},\ and\ \bibinfo {author} {\bibfnamefont
  {J.}~\bibnamefont {Kube{\v{c}}ek}},\ }\bibfield  {title} {\bibinfo {title}
  {How comfortable are your cycling tracks? a new method for objective bicycle
  vibration measurement},\ }\href {https://doi.org/10.1016/j.trc.2015.05.007}
  {\bibfield  {journal} {\bibinfo  {journal} {Transp. Res. C}\ }\textbf
  {\bibinfo {volume} {56}},\ \bibinfo {pages} {415} (\bibinfo {year}
  {2015})}\BibitemShut {NoStop}%
\bibitem [{\citenamefont {Muñoz}\ \emph {et~al.}(2013)\citenamefont {Muñoz},
  \citenamefont {Monzon},\ and\ \citenamefont {Lois}}]{Munoz2013_topo}%
  \BibitemOpen
  \bibfield  {author} {\bibinfo {author} {\bibfnamefont {B.}~\bibnamefont
  {Muñoz}}, \bibinfo {author} {\bibfnamefont {A.}~\bibnamefont {Monzon}},\
  and\ \bibinfo {author} {\bibfnamefont {D.}~\bibnamefont {Lois}},\ }\bibfield
  {title} {\bibinfo {title} {Cycling habits and other psychological variables
  affecting commuting by bicycle in {M}adrid, {S}pain},\ }\href
  {https://doi.org/10.3141/2382-01} {\bibfield  {journal} {\bibinfo  {journal}
  {Transp. Res. Record}\ }\textbf {\bibinfo {volume} {2382}},\ \bibinfo {pages}
  {1} (\bibinfo {year} {2013})}\BibitemShut {NoStop}%
\bibitem [{\citenamefont {Rik~de Groot}(2016)}]{RikdeGroot2016_designmanual}%
  \BibitemOpen
  \bibinfo {editor} {\bibfnamefont {H.}~\bibnamefont {Rik~de Groot}},\ ed.,\
  \href@noop {} {\emph {\bibinfo {title} {Design Manual for Bicycle
  Traffic}}},\ \bibinfo {series} {Record}\ No.~\bibinfo {number} {28}\
  (\bibinfo  {publisher} {{Crow}},\ \bibinfo {year} {2016})\BibitemShut
  {NoStop}%
\bibitem [{\citenamefont {Bushell}\ \emph {et~al.}(2013)\citenamefont
  {Bushell}, \citenamefont {Poole}, \citenamefont {Zegeer},\ and\ \citenamefont
  {Rodriguez}}]{Bushell2013_costs}%
  \BibitemOpen
  \bibfield  {author} {\bibinfo {author} {\bibfnamefont {M.~A.}\ \bibnamefont
  {Bushell}}, \bibinfo {author} {\bibfnamefont {B.~W.}\ \bibnamefont {Poole}},
  \bibinfo {author} {\bibfnamefont {C.~V.}\ \bibnamefont {Zegeer}},\ and\
  \bibinfo {author} {\bibfnamefont {D.~A.}\ \bibnamefont {Rodriguez}},\ }\href
  {https://www.pedbikeinfo.org/cms/downloads/Countermeasure
  Costs_Report_Nov20131.pdf} {\emph {\bibinfo {title} {Costs for Pedestrian and
  Bicyclist Infrastructure Improvements}}},\ \bibinfo {type} {Tech. Rep.}\
  (\bibinfo  {institution} {{UNC Highway Safety Research Center, University of
  North Carolina}},\ \bibinfo {year} {2013})\BibitemShut {NoStop}%
\bibitem [{\citenamefont {Barth{\'e}lemy}(2018)}]{Barthelemy2018}%
  \BibitemOpen
  \bibfield  {author} {\bibinfo {author} {\bibfnamefont {M.}~\bibnamefont
  {Barth{\'e}lemy}},\ }\bibinfo {title} {Optimal networks},\ in\ \href@noop {}
  {\emph {\bibinfo {booktitle} {Morphogenesis of Spatial Networks. Lecture
  Notes in Morphogenesis}}}\ (\bibinfo  {publisher} {Springer International
  Publishing},\ \bibinfo {address} {Cham},\ \bibinfo {year} {2018})\
  Chap.~\bibinfo {chapter} {12}, pp.\ \bibinfo {pages} {241--263}\BibitemShut
  {NoStop}%
\bibitem [{\citenamefont {Natera~Orozco}\ \emph {et~al.}(2020)\citenamefont
  {Natera~Orozco}, \citenamefont {Battiston}, \citenamefont {Iñiguez},\ and\
  \citenamefont {Szell}}]{Natera2020_datadriven}%
  \BibitemOpen
  \bibfield  {author} {\bibinfo {author} {\bibfnamefont {L.~G.}\ \bibnamefont
  {Natera~Orozco}}, \bibinfo {author} {\bibfnamefont {F.}~\bibnamefont
  {Battiston}}, \bibinfo {author} {\bibfnamefont {G.}~\bibnamefont
  {Iñiguez}},\ and\ \bibinfo {author} {\bibfnamefont {M.}~\bibnamefont
  {Szell}},\ }\bibfield  {title} {\bibinfo {title} {Data-driven strategies for
  optimal bicycle network growth},\ }\href
  {https://doi.org/10.1098/rsos.201130} {\bibfield  {journal} {\bibinfo
  {journal} {R. Soc. Open Sci.}\ }\textbf {\bibinfo {volume} {7}},\ \bibinfo
  {pages} {201130} (\bibinfo {year} {2020})}\BibitemShut {NoStop}%
\bibitem [{\citenamefont {Olmos}\ \emph {et~al.}(2020)\citenamefont {Olmos},
  \citenamefont {Tadeo}, \citenamefont {Vlachogiannis}, \citenamefont
  {Alhasoun}, \citenamefont {Alegre}, \citenamefont {Ochoa}, \citenamefont
  {Targa},\ and\ \citenamefont {Gonz{\'a}lez}}]{Olmos2020_data}%
  \BibitemOpen
  \bibfield  {author} {\bibinfo {author} {\bibfnamefont {L.~E.}\ \bibnamefont
  {Olmos}}, \bibinfo {author} {\bibfnamefont {M.~S.}\ \bibnamefont {Tadeo}},
  \bibinfo {author} {\bibfnamefont {D.}~\bibnamefont {Vlachogiannis}}, \bibinfo
  {author} {\bibfnamefont {F.}~\bibnamefont {Alhasoun}}, \bibinfo {author}
  {\bibfnamefont {X.~E.}\ \bibnamefont {Alegre}}, \bibinfo {author}
  {\bibfnamefont {C.}~\bibnamefont {Ochoa}}, \bibinfo {author} {\bibfnamefont
  {F.}~\bibnamefont {Targa}},\ and\ \bibinfo {author} {\bibfnamefont {M.~C.}\
  \bibnamefont {Gonz{\'a}lez}},\ }\bibfield  {title} {\bibinfo {title} {A data
  science framework for planning the growth of bicycle infrastructures},\
  }\href {https://doi.org/10.1016/j.trc.2020.102640} {\bibfield  {journal}
  {\bibinfo  {journal} {Transp. Res. C}\ }\textbf {\bibinfo {volume} {115}},\
  \bibinfo {pages} {102640} (\bibinfo {year} {2020})}\BibitemShut {NoStop}%
\bibitem [{\citenamefont {Szell}\ \emph {et~al.}(2022)\citenamefont {Szell},
  \citenamefont {Mimar}, \citenamefont {Perlman}, \citenamefont {Ghoshal},\
  and\ \citenamefont {Sinatra}}]{Szell2021_growing}%
  \BibitemOpen
  \bibfield  {author} {\bibinfo {author} {\bibfnamefont {M.}~\bibnamefont
  {Szell}}, \bibinfo {author} {\bibfnamefont {S.}~\bibnamefont {Mimar}},
  \bibinfo {author} {\bibfnamefont {T.}~\bibnamefont {Perlman}}, \bibinfo
  {author} {\bibfnamefont {G.}~\bibnamefont {Ghoshal}},\ and\ \bibinfo {author}
  {\bibfnamefont {R.}~\bibnamefont {Sinatra}},\ }\bibfield  {title} {\bibinfo
  {title} {Growing urban bicycle networks},\ }\href
  {https://doi.org/10.1038/s41598-022-10783-y} {\bibfield  {journal} {\bibinfo
  {journal} {Sci. Rep.}\ }\textbf {\bibinfo {volume} {12}},\ \bibinfo {pages}
  {6765} (\bibinfo {year} {2022})}\BibitemShut {NoStop}%
\bibitem [{\citenamefont {Menghini}\ \emph {et~al.}(2010)\citenamefont
  {Menghini}, \citenamefont {Carrasco}, \citenamefont {Schüssler},\ and\
  \citenamefont {Axhausen}}]{Menghini2010_routechoice}%
  \BibitemOpen
  \bibfield  {author} {\bibinfo {author} {\bibfnamefont {G.}~\bibnamefont
  {Menghini}}, \bibinfo {author} {\bibfnamefont {N.}~\bibnamefont {Carrasco}},
  \bibinfo {author} {\bibfnamefont {N.}~\bibnamefont {Schüssler}},\ and\
  \bibinfo {author} {\bibfnamefont {K.}~\bibnamefont {Axhausen}},\ }\bibfield
  {title} {\bibinfo {title} {Route choice of cyclists in Zurich},\ }\href
  {https://doi.org/10.1016/j.tra.2010.07.008} {\bibfield  {journal} {\bibinfo
  {journal} {Transp. Res. A}\ }\textbf {\bibinfo {volume} {44}},\ \bibinfo
  {pages} {754 } (\bibinfo {year} {2010})}\BibitemShut {NoStop}%
\bibitem [{\citenamefont {Broach}\ \emph {et~al.}(2012)\citenamefont {Broach},
  \citenamefont {Dill},\ and\ \citenamefont {Gliebe}}]{Broach2012_where}%
  \BibitemOpen
  \bibfield  {author} {\bibinfo {author} {\bibfnamefont {J.}~\bibnamefont
  {Broach}}, \bibinfo {author} {\bibfnamefont {J.}~\bibnamefont {Dill}},\ and\
  \bibinfo {author} {\bibfnamefont {J.}~\bibnamefont {Gliebe}},\ }\bibfield
  {title} {\bibinfo {title} {Where do cyclists ride? a route choice model
  developed with revealed preference gps data},\ }\href
  {https://doi.org/10.1016/j.tra.2012.07.005} {\bibfield  {journal} {\bibinfo
  {journal} {Transp. Res. A}\ }\textbf {\bibinfo {volume} {46}},\ \bibinfo
  {pages} {1730 } (\bibinfo {year} {2012})}\BibitemShut {NoStop}%
\bibitem [{\citenamefont {Guerreiro}\ \emph {et~al.}(2018)\citenamefont
  {Guerreiro}, \citenamefont {Kirner~Providelo}, \citenamefont {Pitombo},
  \citenamefont {Antonio Rodrigues~Ramos},\ and\ \citenamefont {Rodrigues~da
  Silva}}]{Guerreiro2018_data}%
  \BibitemOpen
  \bibfield  {author} {\bibinfo {author} {\bibfnamefont {T.~d. C.~M.}\
  \bibnamefont {Guerreiro}}, \bibinfo {author} {\bibfnamefont {J.}~\bibnamefont
  {Kirner~Providelo}}, \bibinfo {author} {\bibfnamefont {C.~S.}\ \bibnamefont
  {Pitombo}}, \bibinfo {author} {\bibfnamefont {R.}~\bibnamefont {Antonio
  Rodrigues~Ramos}},\ and\ \bibinfo {author} {\bibfnamefont {A.~N.}\
  \bibnamefont {Rodrigues~da Silva}},\ }\bibfield  {title} {\bibinfo {title}
  {Data-mining, gis and multicriteria analysis in a comprehensive method for
  bicycle network planning and design},\ }\href
  {https://doi.org/10.1080/15568318.2017.1342156} {\bibfield  {journal}
  {\bibinfo  {journal} {Int. J. Sustain. Transp.}\ }\textbf {\bibinfo {volume}
  {12}},\ \bibinfo {pages} {179} (\bibinfo {year} {2018})}\BibitemShut
  {NoStop}%
\bibitem [{\citenamefont {Banister}\ and\ \citenamefont
  {Berechman}(2001)}]{Banister2001}%
  \BibitemOpen
  \bibfield  {author} {\bibinfo {author} {\bibfnamefont {D.}~\bibnamefont
  {Banister}}\ and\ \bibinfo {author} {\bibfnamefont {Y.}~\bibnamefont
  {Berechman}},\ }\bibfield  {title} {\bibinfo {title} {Transport investment
  and the promotion of economic growth},\ }\href
  {https://doi.org/10.1016/S0966-6923(01)00013-8} {\bibfield  {journal}
  {\bibinfo  {journal} {J. Transp. Geogr.}\ }\textbf {\bibinfo {volume} {9}},\
  \bibinfo {pages} {209} (\bibinfo {year} {2001})}\BibitemShut {NoStop}%
\bibitem [{\citenamefont {Buehler}\ and\ \citenamefont
  {Dill}(2016)}]{Buehler2016}%
  \BibitemOpen
  \bibfield  {author} {\bibinfo {author} {\bibfnamefont {R.}~\bibnamefont
  {Buehler}}\ and\ \bibinfo {author} {\bibfnamefont {J.}~\bibnamefont {Dill}},\
  }\bibfield  {title} {\bibinfo {title} {Bikeway networks: a review of effects
  on cycling},\ }\href {https://doi.org/10.1080/01441647.2015.1069908}
  {\bibfield  {journal} {\bibinfo  {journal} {Transp. Rev.}\ }\textbf {\bibinfo
  {volume} {36}},\ \bibinfo {pages} {9} (\bibinfo {year} {2016})}\BibitemShut
  {NoStop}%
\bibitem [{\citenamefont {Caulfield}\ \emph {et~al.}(2012)\citenamefont
  {Caulfield}, \citenamefont {Brick},\ and\ \citenamefont
  {McCarthy}}]{Caulfield2012}%
  \BibitemOpen
  \bibfield  {author} {\bibinfo {author} {\bibfnamefont {B.}~\bibnamefont
  {Caulfield}}, \bibinfo {author} {\bibfnamefont {E.}~\bibnamefont {Brick}},\
  and\ \bibinfo {author} {\bibfnamefont {O.~T.}\ \bibnamefont {McCarthy}},\
  }\bibfield  {title} {\bibinfo {title} {Determining bicycle infrastructure
  preferences – a case study of dublin},\ }\href
  {https://doi.org/10.1016/j.trd.2012.04.001} {\bibfield  {journal} {\bibinfo
  {journal} {Transp. Res. D}\ }\textbf {\bibinfo {volume} {17}},\ \bibinfo
  {pages} {413} (\bibinfo {year} {2012})}\BibitemShut {NoStop}%
\bibitem [{\citenamefont {Wardrop}(1952)}]{Wardrop1952}%
  \BibitemOpen
  \bibfield  {author} {\bibinfo {author} {\bibfnamefont {J.~G.}\ \bibnamefont
  {Wardrop}},\ }\bibfield  {title} {\bibinfo {title} {Road paper. some
  theoretical aspects of road traffic research.},\ }\href
  {https://doi.org/10.1680/ipeds.1952.11259} {\bibfield  {journal} {\bibinfo
  {journal} {Proc. Inst. Civ. Eng.}\ }\textbf {\bibinfo {volume} {1}},\
  \bibinfo {pages} {325} (\bibinfo {year} {1952})}\BibitemShut {NoStop}%
\bibitem [{\citenamefont {Youn}\ \emph {et~al.}(2008)\citenamefont {Youn},
  \citenamefont {Gastner},\ and\ \citenamefont {Jeong}}]{Youn2008_anarchy}%
  \BibitemOpen
  \bibfield  {author} {\bibinfo {author} {\bibfnamefont {H.}~\bibnamefont
  {Youn}}, \bibinfo {author} {\bibfnamefont {M.~T.}\ \bibnamefont {Gastner}},\
  and\ \bibinfo {author} {\bibfnamefont {H.}~\bibnamefont {Jeong}},\ }\bibfield
   {title} {\bibinfo {title} {Price of anarchy in transportation networks:
  Efficiency and optimality control},\ }\href
  {https://doi.org/10.1103/PhysRevLett.101.128701} {\bibfield  {journal}
  {\bibinfo  {journal} {Phys. Rev. Lett.}\ }\textbf {\bibinfo {volume} {101}},\
  \bibinfo {pages} {128701} (\bibinfo {year} {2008})}\BibitemShut {NoStop}%
\bibitem [{\citenamefont {Paulsen}\ and\ \citenamefont
  {Nagel}(2019)}]{Paulsen2019_bikecongestion}%
  \BibitemOpen
  \bibfield  {author} {\bibinfo {author} {\bibfnamefont {M.}~\bibnamefont
  {Paulsen}}\ and\ \bibinfo {author} {\bibfnamefont {K.}~\bibnamefont
  {Nagel}},\ }\bibfield  {title} {\bibinfo {title} {Large-scale assignment of
  congested bicycle traffic using speed heterogeneous agents},\ }\href
  {https://doi.org/10.1016/j.procs.2019.04.112} {\bibfield  {journal} {\bibinfo
   {journal} {Procedia Comput. Sci.}\ }\bibinfo {series} {The 10th
  {{International Conference}} on {{Ambient Systems}}, {{Networks}} and
  {{Technologies}} ({{ANT}} 2019) / {{The}} 2nd {{International Conference}} on
  {{Emerging Data}} and {{Industry}} 4.0 ({{EDI40}} 2019) / {{Affiliated
  Workshops}}},\ \textbf {\bibinfo {volume} {151}},\ \bibinfo {pages} {820}
  (\bibinfo {year} {2019})}\BibitemShut {NoStop}%
\bibitem [{\citenamefont {Daganzo}\ and\ \citenamefont
  {Sheffi}(1977)}]{Daganzo1977}%
  \BibitemOpen
  \bibfield  {author} {\bibinfo {author} {\bibfnamefont {C.~F.}\ \bibnamefont
  {Daganzo}}\ and\ \bibinfo {author} {\bibfnamefont {Y.}~\bibnamefont
  {Sheffi}},\ }\bibfield  {title} {\bibinfo {title} {On stochastic models of
  traffic assignment},\ }\href {https://doi.org/10.1287/trsc.11.3.253}
  {\bibfield  {journal} {\bibinfo  {journal} {Transp. Sci.}\ }\textbf {\bibinfo
  {volume} {11}},\ \bibinfo {pages} {253} (\bibinfo {year} {1977})}\BibitemShut
  {NoStop}%
\bibitem [{\citenamefont {McNeil}\ \emph {et~al.}(2015)\citenamefont {McNeil},
  \citenamefont {Monsere},\ and\ \citenamefont {Dill}}]{McNeil2015}%
  \BibitemOpen
  \bibfield  {author} {\bibinfo {author} {\bibfnamefont {N.}~\bibnamefont
  {McNeil}}, \bibinfo {author} {\bibfnamefont {C.~M.}\ \bibnamefont
  {Monsere}},\ and\ \bibinfo {author} {\bibfnamefont {J.}~\bibnamefont
  {Dill}},\ }\bibfield  {title} {\bibinfo {title} {Influence of bike lane
  buffer types on perceived comfort and safety of bicyclists and potential
  bicyclists},\ }\href {https://doi.org/10.3141/2520-15} {\bibfield  {journal}
  {\bibinfo  {journal} {Transp. Res. Rec.}\ }\textbf {\bibinfo {volume}
  {2520}},\ \bibinfo {pages} {132} (\bibinfo {year} {2015})}\BibitemShut
  {NoStop}%
\bibitem [{\citenamefont {von Stülpnagel}\ and\ \citenamefont
  {Binnig}(2022)}]{Stuelpnagel2022_howsafe}%
  \BibitemOpen
  \bibfield  {author} {\bibinfo {author} {\bibfnamefont {R.}~\bibnamefont {von
  Stülpnagel}}\ and\ \bibinfo {author} {\bibfnamefont {N.}~\bibnamefont
  {Binnig}},\ }\bibfield  {title} {\bibinfo {title} {How safe do you feel?
  {\textendash} a large-scale survey concerning the subjective safety
  associated with different kinds of cycling lanes},\ }\href
  {https://doi.org/10.1016/j.aap.2022.106577} {\bibfield  {journal} {\bibinfo
  {journal} {Accid. Anal. Prev.}\ }\textbf {\bibinfo {volume} {167}},\ \bibinfo
  {pages} {106577} (\bibinfo {year} {2022})}\BibitemShut {NoStop}%
\bibitem [{\citenamefont {Cominetti}\ \emph {et~al.}(2010)\citenamefont
  {Cominetti}, \citenamefont {Melo},\ and\ \citenamefont
  {Sorin}}]{Cominetti2010}%
  \BibitemOpen
  \bibfield  {author} {\bibinfo {author} {\bibfnamefont {R.}~\bibnamefont
  {Cominetti}}, \bibinfo {author} {\bibfnamefont {E.}~\bibnamefont {Melo}},\
  and\ \bibinfo {author} {\bibfnamefont {S.}~\bibnamefont {Sorin}},\ }\bibfield
   {title} {\bibinfo {title} {A payoff-based learning procedure and its
  application to traffic games},\ }\href
  {https://doi.org/10.1016/j.geb.2008.11.012} {\bibfield  {journal} {\bibinfo
  {journal} {Game Econ. Behav.}\ }\textbf {\bibinfo {volume} {70}},\ \bibinfo
  {pages} {71} (\bibinfo {year} {2010})}\BibitemShut {NoStop}%
\bibitem [{\citenamefont {Storch}\ \emph {et~al.}(2020)\citenamefont {Storch},
  \citenamefont {Schr{\"{o}}der},\ and\ \citenamefont {Timme}}]{Storch2020}%
  \BibitemOpen
  \bibfield  {author} {\bibinfo {author} {\bibfnamefont {D.-M.}\ \bibnamefont
  {Storch}}, \bibinfo {author} {\bibfnamefont {M.}~\bibnamefont
  {Schr{\"{o}}der}},\ and\ \bibinfo {author} {\bibfnamefont {M.}~\bibnamefont
  {Timme}},\ }\bibfield  {title} {\bibinfo {title} {Traffic flow splitting from
  crowdsourced digital route choice support},\ }\href
  {https://doi.org/10.1088/2632-072X/aba83e} {\bibfield  {journal} {\bibinfo
  {journal} {J. Phys. Complexity}\ }\textbf {\bibinfo {volume} {1}},\ \bibinfo
  {pages} {035004} (\bibinfo {year} {2020})}\BibitemShut {NoStop}%
\bibitem [{\citenamefont {Newman}\ and\ \citenamefont
  {Girvan}(2004)}]{Newman2004_finding}%
  \BibitemOpen
  \bibfield  {author} {\bibinfo {author} {\bibfnamefont {M.~E.~J.}\
  \bibnamefont {Newman}}\ and\ \bibinfo {author} {\bibfnamefont
  {M.}~\bibnamefont {Girvan}},\ }\bibfield  {title} {\bibinfo {title} {Finding
  and evaluating community structure in networks},\ }\href
  {https://doi.org/10.1103/PhysRevE.69.026113} {\bibfield  {journal} {\bibinfo
  {journal} {Phys. Rev. E}\ }\textbf {\bibinfo {volume} {69}},\ \bibinfo
  {pages} {026113} (\bibinfo {year} {2004})}\BibitemShut {NoStop}%
\bibitem [{\citenamefont {{OpenStreetMap contributors}}(2021)}]{OSM}%
  \BibitemOpen
  \bibfield  {author} {\bibinfo {author} {\bibnamefont {{OpenStreetMap
  contributors}}},\ }\href@noop {} {\bibinfo {title} {Street networks retrieved
  from openstreetmap.org (accessed on 2021-12-02)}} (\bibinfo {year} {2021}),\
  \bibinfo {note} {license: ODbL
  \url{www.openstreetmap.org/copyright/en}}\BibitemShut {NoStop}%
\bibitem [{\citenamefont {nextbike GmbH}(2020)}]{Nextbike}%
  \BibitemOpen
  \bibfield  {author} {\bibinfo {author} {\bibnamefont {nextbike GmbH}},\
  }\href@noop {} {\bibinfo {title} {Trip data {D}resden universities}}
  (\bibinfo {year} {2020}),\ \bibinfo {note} {(data effective
  04/2020)}\BibitemShut {NoStop}%
\bibitem [{\citenamefont {{Deutsche Bahn AG}}(2017)}]{DBData}%
  \BibitemOpen
  \bibfield  {author} {\bibinfo {author} {\bibnamefont {{Deutsche Bahn AG}}},\
  }\href@noop {} {\bibinfo {title} {{B}uchungen {C}all a {B}ike ({S}tand
  05/2017)}} (\bibinfo {year} {2017}),\ \bibinfo {note} {(CC BY 4.0).
  \url{https://data.deutschebahn.com/dataset/data-call-a-bike} (accessed on
  2020-04-14)}\BibitemShut {NoStop}%
\bibitem [{\citenamefont {Kellstedt}\ \emph {et~al.}(2021)\citenamefont
  {Kellstedt}, \citenamefont {Spengler}, \citenamefont {Foster}, \citenamefont
  {Lee},\ and\ \citenamefont {Maddock}}]{Kellstedt2021_bikeability}%
  \BibitemOpen
  \bibfield  {author} {\bibinfo {author} {\bibfnamefont {D.~K.}\ \bibnamefont
  {Kellstedt}}, \bibinfo {author} {\bibfnamefont {J.~O.}\ \bibnamefont
  {Spengler}}, \bibinfo {author} {\bibfnamefont {M.}~\bibnamefont {Foster}},
  \bibinfo {author} {\bibfnamefont {C.}~\bibnamefont {Lee}},\ and\ \bibinfo
  {author} {\bibfnamefont {J.~E.}\ \bibnamefont {Maddock}},\ }\bibfield
  {title} {\bibinfo {title} {A scoping review of bikeability assessment
  methods},\ }\href {https://doi.org/10.1007/s10900-020-00846-4} {\bibfield
  {journal} {\bibinfo  {journal} {J. Community Health}\ }\textbf {\bibinfo
  {volume} {46}},\ \bibinfo {pages} {211} (\bibinfo {year} {2021})}\BibitemShut
  {NoStop}%
\bibitem [{\citenamefont {Gebhart}\ and\ \citenamefont
  {Noland}(2014)}]{Gebhart2014_weather}%
  \BibitemOpen
  \bibfield  {author} {\bibinfo {author} {\bibfnamefont {K.}~\bibnamefont
  {Gebhart}}\ and\ \bibinfo {author} {\bibfnamefont {R.~B.}\ \bibnamefont
  {Noland}},\ }\bibfield  {title} {\bibinfo {title} {The impact of weather
  conditions on bikeshare trips in {W}ashington, {DC}},\ }\href
  {https://doi.org/10.1007/s11116-014-9540-7} {\bibfield  {journal} {\bibinfo
  {journal} {Transportation}\ }\textbf {\bibinfo {volume} {41}},\ \bibinfo
  {pages} {1205} (\bibinfo {year} {2014})}\BibitemShut {NoStop}%
\bibitem [{\citenamefont {Saneinejad}\ \emph {et~al.}(2012)\citenamefont
  {Saneinejad}, \citenamefont {Roorda},\ and\ \citenamefont
  {Kennedy}}]{Saneinejad2012_weather}%
  \BibitemOpen
  \bibfield  {author} {\bibinfo {author} {\bibfnamefont {S.}~\bibnamefont
  {Saneinejad}}, \bibinfo {author} {\bibfnamefont {M.~J.}\ \bibnamefont
  {Roorda}},\ and\ \bibinfo {author} {\bibfnamefont {C.}~\bibnamefont
  {Kennedy}},\ }\bibfield  {title} {\bibinfo {title} {Modelling the impact of
  weather conditions on active transportation travel behaviour},\ }\href
  {https://doi.org/10.1016/j.trd.2011.09.005} {\bibfield  {journal} {\bibinfo
  {journal} {Transp. Res. D}\ }\textbf {\bibinfo {volume} {17}},\ \bibinfo
  {pages} {129 } (\bibinfo {year} {2012})}\BibitemShut {NoStop}%
\bibitem [{\citenamefont {Buehler}\ and\ \citenamefont
  {Pucher}(2011)}]{Buehler2011_impact_bikepaths}%
  \BibitemOpen
  \bibfield  {author} {\bibinfo {author} {\bibfnamefont {R.}~\bibnamefont
  {Buehler}}\ and\ \bibinfo {author} {\bibfnamefont {J.}~\bibnamefont
  {Pucher}},\ }\bibfield  {title} {\bibinfo {title} {Cycling to work in 90
  large american cities: new evidence on the role of bike paths and lanes},\
  }\href {https://doi.org/10.1007/s11116-011-9355-8} {\bibfield  {journal}
  {\bibinfo  {journal} {Transportation}\ }\textbf {\bibinfo {volume} {39}},\
  \bibinfo {pages} {409} (\bibinfo {year} {2011})}\BibitemShut {NoStop}%
\bibitem [{\citenamefont {Carmona}\ \emph {et~al.}(2020)\citenamefont
  {Carmona}, \citenamefont {de~Noronha}, \citenamefont {Moreira}, \citenamefont
  {Ara\'ujo},\ and\ \citenamefont {Andrade}}]{Carmona2020_cracking}%
  \BibitemOpen
  \bibfield  {author} {\bibinfo {author} {\bibfnamefont {H.~A.}\ \bibnamefont
  {Carmona}}, \bibinfo {author} {\bibfnamefont {A.~W.~T.}\ \bibnamefont
  {de~Noronha}}, \bibinfo {author} {\bibfnamefont {A.~A.}\ \bibnamefont
  {Moreira}}, \bibinfo {author} {\bibfnamefont {N.~A.~M.}\ \bibnamefont
  {Ara\'ujo}},\ and\ \bibinfo {author} {\bibfnamefont {J.~S.}\ \bibnamefont
  {Andrade}},\ }\bibfield  {title} {\bibinfo {title} {Cracking urban
  mobility},\ }\href {https://doi.org/10.1103/PhysRevResearch.2.043132}
  {\bibfield  {journal} {\bibinfo  {journal} {Phys. Rev. Research}\ }\textbf
  {\bibinfo {volume} {2}},\ \bibinfo {pages} {043132} (\bibinfo {year}
  {2020})}\BibitemShut {NoStop}%
\bibitem [{\citenamefont {Schäfer}\ \emph {et~al.}(2018)\citenamefont
  {Schäfer}, \citenamefont {Witthaut}, \citenamefont {Timme},\ and\
  \citenamefont {Latora}}]{Schafer2018_Cascading}%
  \BibitemOpen
  \bibfield  {author} {\bibinfo {author} {\bibfnamefont {B.}~\bibnamefont
  {Schäfer}}, \bibinfo {author} {\bibfnamefont {D.}~\bibnamefont {Witthaut}},
  \bibinfo {author} {\bibfnamefont {M.}~\bibnamefont {Timme}},\ and\ \bibinfo
  {author} {\bibfnamefont {V.}~\bibnamefont {Latora}},\ }\bibfield  {title}
  {\bibinfo {title} {Dynamically induced cascading failures in power grids},\
  }\href {https://doi.org/10.1038/s41467-018-04287-5} {\bibfield  {journal}
  {\bibinfo  {journal} {Nat. Commun.}\ }\textbf {\bibinfo {volume} {9}},\
  \bibinfo {pages} {1975} (\bibinfo {year} {2018})}\BibitemShut {NoStop}%
\bibitem [{\citenamefont {Kirkpatrick}\ \emph {et~al.}(1983)\citenamefont
  {Kirkpatrick}, \citenamefont {Gelatt},\ and\ \citenamefont
  {Vecchi}}]{Kirkpatrick671_Annealing}%
  \BibitemOpen
  \bibfield  {author} {\bibinfo {author} {\bibfnamefont {S.}~\bibnamefont
  {Kirkpatrick}}, \bibinfo {author} {\bibfnamefont {C.~D.}\ \bibnamefont
  {Gelatt}},\ and\ \bibinfo {author} {\bibfnamefont {M.~P.}\ \bibnamefont
  {Vecchi}},\ }\bibfield  {title} {\bibinfo {title} {Optimization by simulated
  annealing},\ }\href {https://doi.org/10.1126/science.220.4598.671} {\bibfield
   {journal} {\bibinfo  {journal} {Science}\ }\textbf {\bibinfo {volume}
  {220}},\ \bibinfo {pages} {671} (\bibinfo {year} {1983})}\BibitemShut
  {NoStop}%
\bibitem [{\citenamefont {Kirkegaard}\ and\ \citenamefont
  {Sneppen}(2020)}]{Kirkegaard2020_annealing}%
  \BibitemOpen
  \bibfield  {author} {\bibinfo {author} {\bibfnamefont {J.~B.}\ \bibnamefont
  {Kirkegaard}}\ and\ \bibinfo {author} {\bibfnamefont {K.}~\bibnamefont
  {Sneppen}},\ }\bibfield  {title} {\bibinfo {title} {Optimal transport flows
  for distributed production networks},\ }\href
  {https://doi.org/10.1103/PhysRevLett.124.208101} {\bibfield  {journal}
  {\bibinfo  {journal} {Phys. Rev. Lett.}\ }\textbf {\bibinfo {volume} {124}},\
  \bibinfo {pages} {208101} (\bibinfo {year} {2020})}\BibitemShut {NoStop}%
\bibitem [{\citenamefont {{OECD}}(2021)}]{OECD2021}%
  \BibitemOpen
  \bibfield  {author} {\bibinfo {author} {\bibnamefont {{OECD}}},\ }\href
  {https://doi.org/10.1787/0a20f779-en} {\emph {\bibinfo {title} {Transport
  Strategies for Net-Zero Systems by Design}}}\ (\bibinfo  {publisher}
  {{OECD}},\ \bibinfo {year} {2021})\BibitemShut {NoStop}%
\bibitem [{\citenamefont {Folco}\ \emph {et~al.}(2022)\citenamefont {Folco},
  \citenamefont {Gauvin}, \citenamefont {Tizzoni},\ and\ \citenamefont
  {Szell}}]{Folco2022_safety}%
  \BibitemOpen
  \bibfield  {author} {\bibinfo {author} {\bibfnamefont {P.}~\bibnamefont
  {Folco}}, \bibinfo {author} {\bibfnamefont {L.}~\bibnamefont {Gauvin}},
  \bibinfo {author} {\bibfnamefont {M.}~\bibnamefont {Tizzoni}},\ and\ \bibinfo
  {author} {\bibfnamefont {M.}~\bibnamefont {Szell}},\ }\bibfield  {title}
  {\bibinfo {title} {Data-driven bicycle network planning for demand and
  safety},\ }\href@noop {} {\bibfield  {journal} {\bibinfo  {journal} {arXiv
  preprint}\ } (\bibinfo {year} {2022})},\ \Eprint
  {https://arxiv.org/abs/2203.14619} {arXiv:2203.14619} \BibitemShut {NoStop}%
\bibitem [{\citenamefont {Boeing}(2017)}]{OSMnx}%
  \BibitemOpen
  \bibfield  {author} {\bibinfo {author} {\bibfnamefont {G.}~\bibnamefont
  {Boeing}},\ }\bibfield  {title} {\bibinfo {title} {{OSMnx}: New methods for
  acquiring, constructing, analyzing, and visualizing complex street
  networks},\ }\href {https://doi.org/10.1016/j.compenvurbsys.2017.05.004}
  {\bibfield  {journal} {\bibinfo  {journal} {Comput. Environ. Urban Syst.}\
  }\textbf {\bibinfo {volume} {65}},\ \bibinfo {pages} {126} (\bibinfo {year}
  {2017})}\BibitemShut {NoStop}%
\bibitem [{\citenamefont {Graser}\ \emph {et~al.}(2014)\citenamefont {Graser},
  \citenamefont {Straub},\ and\ \citenamefont
  {Dragaschnig}}]{Graser2014_osmquality}%
  \BibitemOpen
  \bibfield  {author} {\bibinfo {author} {\bibfnamefont {A.}~\bibnamefont
  {Graser}}, \bibinfo {author} {\bibfnamefont {M.}~\bibnamefont {Straub}},\
  and\ \bibinfo {author} {\bibfnamefont {M.}~\bibnamefont {Dragaschnig}},\
  }\bibfield  {title} {\bibinfo {title} {Towards an open source analysis
  toolbox for street network comparison: Indicators, tools and results of a
  comparison of osm and the official austrian reference graph},\ }\href
  {https://doi.org/10.1111/tgis.12061} {\bibfield  {journal} {\bibinfo
  {journal} {Trans. GIS}\ }\textbf {\bibinfo {volume} {18}},\ \bibinfo
  {pages} {510} (\bibinfo {year} {2014})}\BibitemShut {NoStop}%
\bibitem [{\citenamefont {Quinn}\ and\ \citenamefont
  {Bull}(2019)}]{Quinn2019_osmquality}%
  \BibitemOpen
  \bibfield  {author} {\bibinfo {author} {\bibfnamefont {S.}~\bibnamefont
  {Quinn}}\ and\ \bibinfo {author} {\bibfnamefont {F.}~\bibnamefont {Bull}},\
  }\bibinfo {title} {Geospatial information system use in public
  organizations}\ (\bibinfo  {publisher} {Routledge},\ \bibinfo {year} {2019})\
  Chap.\ \bibinfo {chapter} {Understanding Threats to Crowdsourced Geographic
  Data Quality Through a Study of OpenStreetMap Contributor Bans}, pp.\
  \bibinfo {pages} {80--96}\BibitemShut {NoStop}%
\bibitem [{\citenamefont {{OpenStreetMap contributors}}(2022)}]{OSMwiki}%
  \BibitemOpen
  \bibfield  {author} {\bibinfo {author} {\bibnamefont {{OpenStreetMap
  contributors}}},\ }\href@noop {} {\bibinfo {title} {Key:highway --- {OSM
  Wiki}}} (\bibinfo {year} {2022}),\ \bibinfo {note}
  {\url{https://wiki.openstreetmap.org/wiki/Key:highway} (accessed on
  2020-12-02)}\BibitemShut {NoStop}%
\bibitem [{\citenamefont {Steinacker}(2022)}]{GitHub_repo}%
  \BibitemOpen
  \bibfield  {author} {\bibinfo {author} {\bibfnamefont {C.}~\bibnamefont
  {Steinacker}},\ }\href {https://doi.org/10.5281/zenodo.6602117} {\bibinfo
  {title} {{BikePathNet}}},\ \bibinfo {howpublished}
  {\url{https://www.doi.org/10.5281/zenodo.6602117}} (\bibinfo {year}
  {2022})\BibitemShut {NoStop}%
\bibitem [{\citenamefont {{Motivate International Inc.}}\ and\ \citenamefont
  {{City of Chicago}}(2018)}]{ChicagoData}%
  \BibitemOpen
  \bibfield  {author} {\bibinfo {author} {\bibnamefont {{Motivate International
  Inc.}}}\ and\ \bibinfo {author} {\bibnamefont {{City of Chicago}}},\ }\href
  {https://www.divvybikes.com/system-data} {\bibinfo {title} {{C}hichago
  {D}ivvy trip data}} (\bibinfo {year} {2018}),\ \bibinfo {note}
  {\url{https://www.divvybikes.com/system-data} (accessed on
  2019-01-09)}\BibitemShut {NoStop}%
\bibitem [{\citenamefont {{Bay Area Motivate, LLC}}(2018)}]{BayAreaData}%
  \BibitemOpen
  \bibfield  {author} {\bibinfo {author} {\bibnamefont {{Bay Area Motivate,
  LLC}}},\ }\href {https://www.lyft.com/bikes/bay-wheels/system-data} {\bibinfo
  {title} {{B}ay {W}heels's trip data}} (\bibinfo {year} {2018}),\ \bibinfo
  {note} {\url{https://www.lyft.com/bikes/bay-wheels/system-data} (accessed on
  2020-03-24)}\BibitemShut {NoStop}%
\bibitem [{\citenamefont {{NYC Bike Share, LLC and Jersey City Bike Share,
  LLC}}(2018)}]{NYCData}%
  \BibitemOpen
  \bibfield  {author} {\bibinfo {author} {\bibnamefont {{NYC Bike Share, LLC
  and Jersey City Bike Share, LLC}}},\ }\href
  {https://www.citibikenyc.com/system-data} {\bibinfo {title} {{C}iti {B}ike
  trip data}} (\bibinfo {year} {2018}),\ \bibinfo {note} {{NYCBS Data Use
  Policy}, \url{https://www.citibikenyc.com/system-data} (accessed on
  2020-01-21)}\BibitemShut {NoStop}%
\bibitem [{\citenamefont {{BIXI Montréal}}(2018)}]{MontrealData}%
  \BibitemOpen
  \bibfield  {author} {\bibinfo {author} {\bibnamefont {{BIXI Montréal}}},\
  }\href {https://www.bixi.com/en/open-data} {\bibinfo {title} {{{BIXI}}
  montréal trip data}} (\bibinfo {year} {2018}),\ \bibinfo {note}
  {\url{https://www.bixi.com/en/open-data} (accessed on
  2020-04-14)}\BibitemShut {NoStop}%
\bibitem [{\citenamefont {{Toronto Parking
  Authority}}(2018)}]{TorontoParkingAuthority}%
  \BibitemOpen
  \bibfield  {author} {\bibinfo {author} {\bibnamefont {{Toronto Parking
  Authority}}},\ }\href
  {https://open.toronto.ca/dataset/bike-share-toronto-ridership-data/}
  {\bibinfo {title} {Bike share {T}oronto ridership data}} (\bibinfo {year}
  {2018}),\ \bibinfo {note} {{Open Government Licence},
  \url{https://open.toronto.ca/dataset/bike-share-toronto-ridership-data/}
  (accessed on 2020-04-20)}\BibitemShut {NoStop}%
\bibitem [{\citenamefont {{Motivate International
  Inc}}(2019)}]{WashingtonData}%
  \BibitemOpen
  \bibfield  {author} {\bibinfo {author} {\bibnamefont {{Motivate International
  Inc}}},\ }\href {https://www.capitalbikeshare.com/system-data} {\bibinfo
  {title} {{C}apital {B}ikeshare trip data}} (\bibinfo {year} {2019}),\
  \bibinfo {note} {{Capital Bikeshare Data License Agreement},
  \url{https://www.capitalbikeshare.com/system-data} (accessed on
  2020-01-21)}\BibitemShut {NoStop}%
\bibitem [{\citenamefont {{Oslo Bysykkel}}(2019)}]{OsloBysykkel}%
  \BibitemOpen
  \bibfield  {author} {\bibinfo {author} {\bibnamefont {{Oslo Bysykkel}}},\
  }\href {https://oslobysykkel.no/en/open-data/historical} {\bibinfo {title}
  {{O}slo {B}ysykkel trip data}} (\bibinfo {year} {2019}),\ \bibinfo {note}
  {{Norsk lisens for offentlige data (NLOD) 2.0},
  \url{https://oslobysykkel.no/en/open-data/historical} (accessed on
  2020-01-01)}\BibitemShut {NoStop}%
\bibitem [{\citenamefont {{Helsinki Region Transport}}(2019)}]{HelsinkiData}%
  \BibitemOpen
  \bibfield  {author} {\bibinfo {author} {\bibnamefont {{Helsinki Region
  Transport}}},\ }\href {https://www.hsl.fi/en/opendata} {\bibinfo {title}
  {{H}elsinki region trip data}} (\bibinfo {year} {2019}),\ \bibinfo {note}
  {{CC BY 4.0}, \url{https://www.hsl.fi/en/opendata} (accessed on
  2020-04-14)}\BibitemShut {NoStop}%
\bibitem [{\citenamefont {{ECOBICI}}\ and\ \citenamefont {{Government of Mexico
  City}}(2019)}]{MexicoCityData}%
  \BibitemOpen
  \bibfield  {author} {\bibinfo {author} {\bibnamefont {{ECOBICI}}}\ and\
  \bibinfo {author} {\bibnamefont {{Government of Mexico City}}},\ }\href
  {https://www.ecobici.cdmx.gob.mx/en/informacion-del-servicio/open-data}
  {\bibinfo {title} {Mexico {C}ity {{ECOBICI}} trip data}} (\bibinfo {year}
  {2019}),\ \bibinfo {note}
  {\url{https://www.ecobici.cdmx.gob.mx/en/informacion-del-servicio/open-data}
  (accessed on 2020-04-14)}\BibitemShut {NoStop}%
\end{thebibliography}
\end{document}